\documentclass[a4paper,fleqn]{cas-sc}

\usepackage[numbers]{natbib}

\def\tsc#1{\csdef{#1}{\textsc{\lowercase{#1}}\xspace}}
\tsc{WGM}
\tsc{QE}
\usepackage{amssymb}
\usepackage{latexsym}
\usepackage{amsmath}

\usepackage{subcaption}
\usepackage{float}

\usepackage{caption}
\usepackage{booktabs}
\usepackage{tabularx}

\usepackage{siunitx}
\usepackage[normalem]{ulem}

\DeclareMathOperator{\grad}{\nabla}
\DeclareMathOperator{\dive}{\nabla\cdot}

\begin{document}
\definecolor{dartmouthgreen}{rgb}{0.05, 0.5, 0.06}	

\let\WriteBookmarks\relax
\def\floatpagepagefraction{1}
\def\textpagefraction{.001}
	
\shorttitle{DG curved boundary for flows over orography}    
	
\shortauthors{G. Orlando et al.}  
	
\title[mode = title]{Impact of curved elements for flows over orography with a Discontinuous Galerkin scheme}  
	
\author[1]{Giuseppe Orlando}[orcid=0000-0002-7119-4231]
\cormark[1]
\ead{giuseppe.orlando@polytechnique.edu}

\author[2]{Tommaso Benacchio}[orcid=0000-0002-0732-7167]
\ead{tbo@dmi.dk}
	
\author[3]{Luca Bonaventura}[orcid=0000-0002-1994-0217]
\ead{luca.bonaventura@polimi.it}
	
\affiliation[1]{organization={CMAP, CNRS, \'{E}cole polytechnique, Institut Polytechnique de Paris},
addressline={Route de Saclay}, 
city={Palaiseau},
postcode={91120}, 
country={France}}

\affiliation[2]{organization={Weather Research, Danish Meteorological Institute},
addressline={Sankt Kjelds Plads 11}, 
city={Copenaghen},
postcode={2100}, 
country={Denmark}}

\affiliation[3]{organization={Dipartimento di Matematica, Politecnico di Milano},
addressline={Piazza Leonardo da Vinci 32}, 
city={Milano},
postcode={20133}, 
country={Italy}}
	
\cortext[1]{Corresponding author}
	
\begin{abstract}
We present a quantitative assessment of the impact of high-order mappings on the simulation of flows over complex orography. Curved boundaries were not used in early numerical methods, whereas they are employed to an increasing extent in state of the art computational fluid dynamics codes, in combination with high-order methods, such as the Finite Element Method and the Spectral Element Method. Here we consider a specific Discontinuous Galerkin (DG) method implemented in the framework of the \texttt{deal.II} library, which natively supports high-order mappings. A number of numerical experiments based on classical benchmarks over idealized orographic profiles demonstrate the positive impact of curved boundaries on the accuracy of the results, with no significantly adverse effect on the computational cost of the simulation. These findings are also supported by results of the application of this approach to non-smooth and realistic orographic profiles.
\end{abstract}
	
	
% Keywords
	
\maketitle

%%%%%%%%%%%%%%%%%%%%%%%%%% Introduction %%%%%%%%%%%%%%%%%%%%%%%%%%%%%%%%%%
\section{Introduction}
\label{sec:intro} \indent

High-order numerical methods are employed to an increasing extent for many relevant physical problems, in particular for applications in Computational Fluid Dynamics (CFD), see e.g. \cite{bassi:1997, dolejsi:2004, gaburro:2023}, among many others. Moreover, in recent years, a general effort to design numerical methods allowing for more general element shapes, such as polygonal/polyhedral elements, is witnessed, see, e.g., \cite{boscheri:2024, corti:2023, fumagalli:2024}, and the references therein. However, such methods must be supplemented by high-order approximations of curved geometries to maintain their high-order accuracy \cite{hindenlang:2015} and achieve optimal convergence rates \cite{ciarlet:2002}. For instance, considering Dirichlet boundary conditions, any finite element method on curved domains is at most second order accurate \cite{strang:1973, thomee:1971}, unless a curved boundary element is adopted \cite{ciarlet:2002}. Thus, the use of curved meshes takes on considerable importance for many realistic applications, and the treatment of boundary conditions for curved boundaries is an active research area, see e.g. the recent contributions \cite{ciallella:2023a, ciallella:2023b}. While the use of numerical schemes in combination with curved boundaries posed limitations for early numerical methods, a number of approaches to deal naturally with curved geometries have been proposed for high-order methods over the last fifty years. 

A very popular approach to build a curvilinear element method relies on the iso-parametric approximation of the curved boundary which, since the seminal contributions \cite{gordon:1973, zienkiewicz:1971}, has widely been employed in the literature, see among many others \cite{dey:2001, fortunato:2016, moxey:2016, toulorge:2013, turner:2018}. Advanced developments use rational B-spline or NURBS approximations in the so-called iso-geometric analysis (IGA) framework \cite{hughes:2005}. An alternative to curvilinear elements is to improve the treatment of the boundary conditions, taking into account the features of the real curved geometry on a straight faced mesh \cite{krivodonova:2006, wang:2003}. However, this method can be formulated only for slip-wall conditions and for 2D geometries. We also mention the approach described, e.g, in \cite{antonietti:2024}, in which a sub-tessellation is considered to evaluate integrals on straight faced meshes. Finally, recent developments employ a reconstruction off-site data approach \cite{costa:2018, costa:2019}, using a least-squares method to handle several constraints imposed by scattered mean values associated to the elements, while yet another approach employs the so-called Shifted Boundary Method, see \cite{ciallella:2023a}, to which we refer for all the details.

We focus here on the Discontinuous Galerkin (DG) method, which combines high-order accuracy and flexibility \cite{giraldo:2020}. A theoretical analysis on the applicability of the DG method on essentially arbitrary shaped-elements, including curved polygonal/polyhedral elements, was presented in \cite{cangiani:2022}. The iso-parametric approximation and IGA have been widely used in conjunction with DG schemes \cite{bassi:1997, hesthaven:2002, kopriva:2009}. However, these approaches are particularly expensive, both in terms of computational cost and memory overhead. A simpler treatment of curved elements in the context of DG methods uses instead high-order mappings from the straight-sided reference element to each curved element \cite{lomtev:1999, wang:2009}, leading to a non-trivial expression for the determinant of the mapping Jacobian (see Section \ref{sec:high_order}). 

While the accuracy loss due to the use of low-order mappings can be clearly identified and analyzed when dealing with smooth boundaries, the use and impact of high-order mappings in presence of irregular and non-smooth boundaries is less straightforward. This aspect is especially important when dealing with applications to numerical weather prediction (NWP) and climate simulations, in which the lower boundary is defined by irregular orographic data. Indeed, in NWP and climate models it is often necessary to filter the orographic data to avoid or reduce spurious numerical effects \cite{webster:2003}. Subgrid-scale orographic drag parametrizations are then employed to compensate the insufficient resolution of orographic features \cite{miller:1989, palmer:1986}. On the other hand, increasing the resolution without parametrization is beneficial to improve the description of atmospheric processes over complex orography and forecast skill \cite{fritts:2022, kanehama:2019}.The importance of an accurate description of orography is also highlighted in studies like \cite{prusa:2006}, where the impact of improved orographic resolution is assessed via idealized climate simulations. Moreover, the positive impact of parametrizations decreases as the model resolution increases. Hence, it is important to assess the performance of high-order mappings also in the case of non-smooth orography and to analyse the interplay of those mappings with standard filtering procedures. Atmospheric motion over idealized orography is the focus of a number of popular benchmarks proposed and analyzed in the NWP literature \cite{bonaventura:2000, klemp:1983, klemp:1978, schar:2002}. The orography is typically described by non-linear smooth analytical profiles and therefore the computational domain is characterized by a curved boundary. Nevertheless, to the best of our knowledge, curved elements have not been employed for such benchmarks and no survey of their impact on numerical results is available. 

In this work, we assess the impact of high-order mappings on the accuracy of numerical simulations of atmospheric flow over both idealized and real orography. For the purpose of this study, we use the IMEX-DG scheme proposed in \cite{orlando:2023b, orlando:2022}, to which we refer for a complete analysis and description of the method. The method was validated for the simulations of weakly compressible atmospheric flows in \cite{orlando:2023a, orlando:2024}. The solver is implemented in the framework of the open-source library \texttt{deal.II} \cite{arndt:2023, bangerth:2007}, which natively supports high-order mappings. We show that, for smooth orography profiles, using high-order mappings leads to more accurate results than using linear mapping. In addition, high-order mappings are found to provide analogous results to those obtained with linear mappings at higher resolution, meaning that the use of high-order mappings acts as a sort of sub-tessellation. Next, we modify the customary smooth benchmarks by adding a non-differentiable perturbation. We compare the results of very high resolution simulations using linear mappings that resolve well the irregular profile to those of lower resolution simulations carried out with both low- and high-order mappings. High-order mappings are found to provide more accurate results also in this case. We then assess the impact of two common filtering procedures on the results. Finally, we consider the use of high-order mappings for a realistic complex orography described by a set of point data.

The paper is structured as follows. The model equations are briefly presented in Section \ref{sec:modeleq}. A short review of the use of high-order mappings for DG schemes is reported in Section \ref{sec:high_order}. Relevant numerical simulations showing the impact of high-order mappings and curved elements are presented in Section \ref{sec:num}. Finally, some conclusions are reported in Section \ref{sec:conclu}. 

%%%%%%%%%%%%%%%%%%%%%%%%%%%% Model equations %%%%%%%%%%%%%%%%%%%%%%%%%
\section{The model equations}
\label{sec:modeleq} \indent

The mathematical model consists of the fully compressible Euler equations of gas dynamics in conservation form under the influence of gravity, completed with the ideal gas law \cite{orlando:2023a, orlando:2024}, in a vertical \((x,z)\) slice domain. Let \(\Omega \subset \mathbb{R}^{d}, d = 2\), be a connected open bounded set with a possibly curved boundary \(\partial\Omega\), with \(\mathbf{x}\) denoting the spatial coordinates and \(t\) the temporal coordinate. We consider the case \(d = 2\) in this work, but the analysis and the considerations outlined in the following sections straightforward extend to \(d = 3\). No \textit{ad hoc} two-dimensional numerical procedure or code was employed for the numerical results presented in this work. We consider the unsteady compressible Euler equations, written in conservation form as
\begin{eqnarray}\label{eq:euler_comp}
	\frac{\partial\rho}{\partial t} + \dive\left(\rho\mathbf{u}\right) &=& 0 \nonumber \\
	\frac{\partial\left(\rho\mathbf{u}\right)}{\partial t} + \dive\left(\rho\mathbf{u} \otimes \mathbf{u}\right) + \grad p &=& \rho\mathbf{g} \\
	\frac{\partial\left(\rho E\right)}{\partial t} + \dive\left[\left(\rho E + p\right)\mathbf{u}\right] &=& \rho \mathbf{g} \cdot \mathbf{u}, \nonumber
\end{eqnarray}
for \(\mathbf{x} \in \Omega\), \(t \in (0, T_{f}]\), endowed with suitable initial and boundary conditions. Here \(T_{f}\) is the final time, \(\rho\) is the density, \(\mathbf{u}\) is the fluid velocity, \(p\) is the pressure, and \(\otimes\) denotes the tensor product. Moreover, \(\mathbf{g} = -g\mathbf{k}\) represents the acceleration of gravity, with \(g = \SI{9.81}{\meter\per\second\squared}\) and \(\mathbf{k}\) denoting the upward pointing unit vector in the standard Cartesian frame of reference. The total energy \(\rho E\) can be rewritten as \(\rho E = \rho e + \rho k\), where \(e\) is the internal energy and \(k = \frac{1}{2}\left|\mathbf{u}\right|^{2}\) is the kinetic energy. We also introduce the specific enthalpy \(h = e + \frac{p}{\rho}\) and we notice that one can rewrite the energy flux as
\begin{equation}
	\left(\rho E + p\right)\mathbf{u} = \left(e + k + \frac{p}{\rho}\right)\rho\mathbf{u} = \left(h + k\right)\rho\mathbf{u}.
\end{equation}
The above equations are complemented by the equation of state for ideal gases, given by 
\begin{equation}
	p = \rho RT,
\end{equation}
with \(R\) being the specific gas constant. For later reference, we define the Exner pressure \(\Pi\) as
\begin{equation}
	\Pi = \left(\frac{p_{0}}{p}\right)^{\frac{\gamma - 1}{\gamma}},
\end{equation}
with \(p_{0} = \SI[parse-numbers=false]{10^{5}}{\pascal}\) being a reference pressure and \(\gamma\) denoting the specific heats ratio. We take the specific heats ratio \(\gamma = 1.4\) and the specific gas constant \(R = \SI{287}{\joule\per\kilo\gram\per\kelvin}\).

%%%%%%%%%%%%%%%%%%%%%%%%% High-order mappings %%%%%%%%%%%%%%%%%%%%%%%%%
\section{High-order mappings for DG schemes}
\label{sec:high_order} \indent

We consider a tessellation of the domain \(\Omega\) into a family of quadrilaterals \(\mathcal{T}_{h}\) and denote each element by \(K\). The most classical approach for the treatment of curved boundaries in the finite element method is the iso-parametric approximation, in which both the geometry and the solution are approximated by some high-order polynomial functions \cite{gordon:1973, zienkiewicz:1971}. In particular, rational B-spline or NURBS approximations lead to the iso-geometric analysis (IGA) \cite{hughes:2005}. Even though its effectiveness has been shown in several contributions, see e.g. \cite{moxey:2016, sahni:2010, toulorge:2013, turner:2018}, this approach can be expensive in terms of computational cost and memory overhead, because of the integration on curvilinear elements, especially when the boundary geometry is represented by very high-order polynomials \cite{yin:2021, zhang:2016}.

\begin{figure}[h!]
	\centering
	\begin{subfigure}{0.475\textwidth}
		\centering
		\includegraphics[width = 0.95\textwidth]{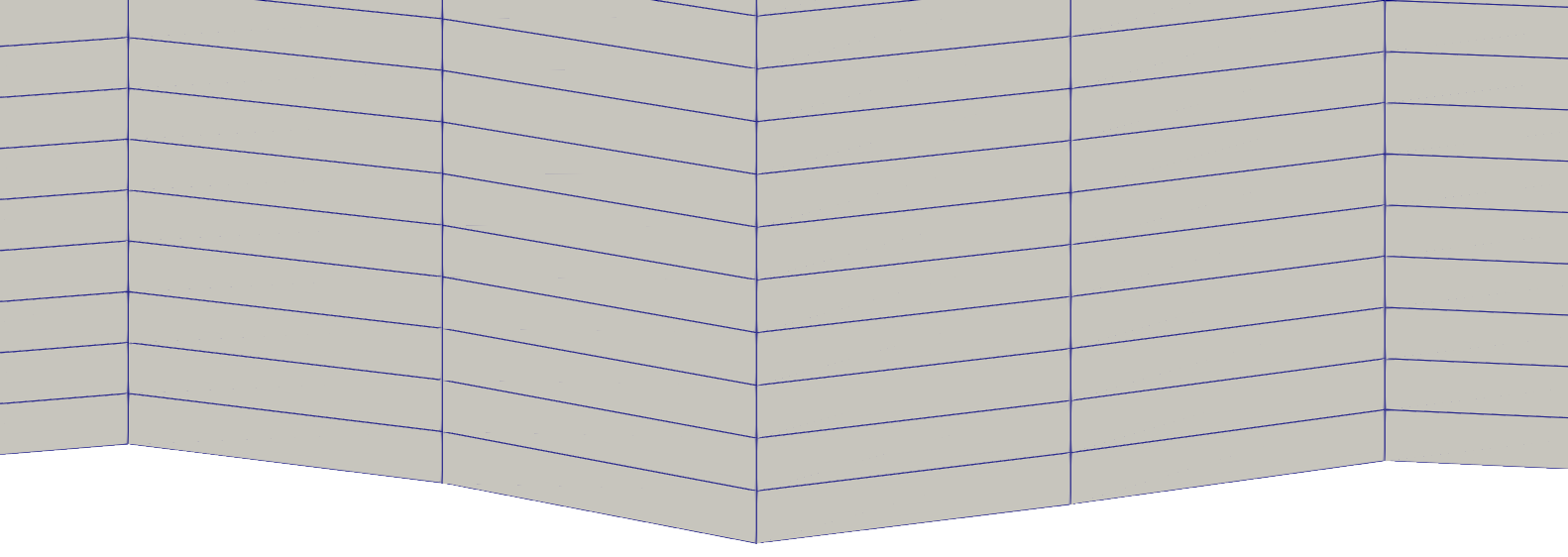}
	\end{subfigure}
	\begin{subfigure}{0.475\textwidth}
		\centering
		\includegraphics[width = 0.95\textwidth]{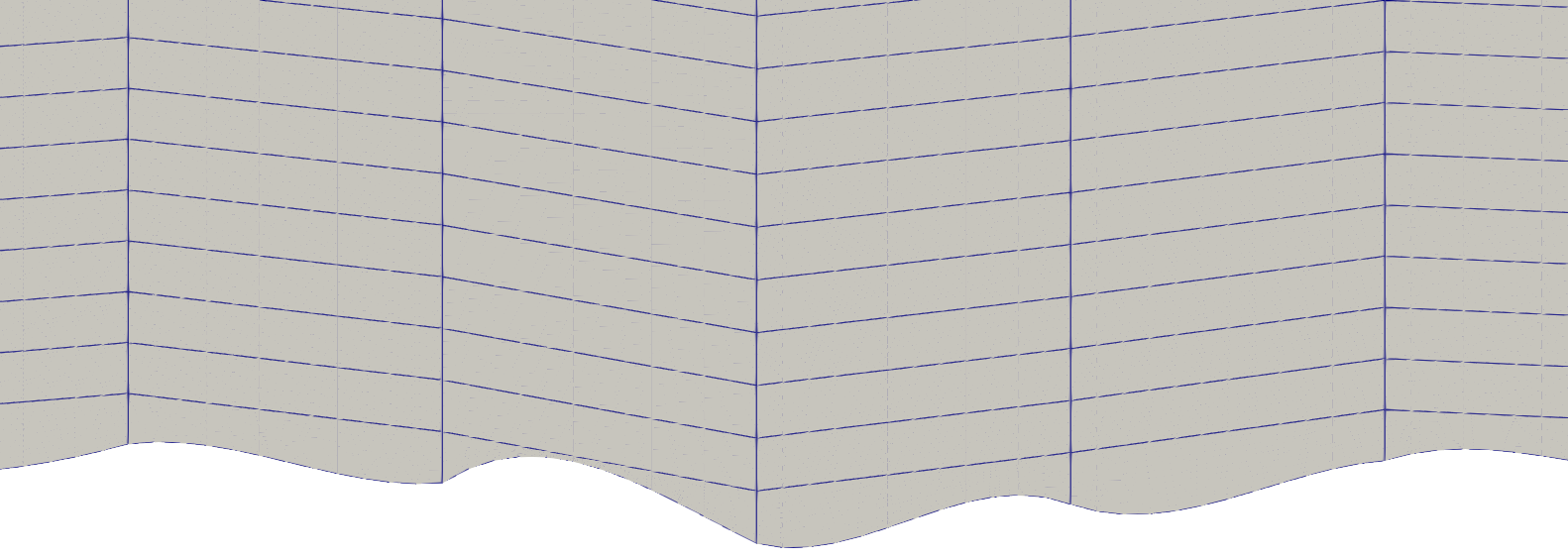}
	\end{subfigure}
	\caption{Comparison between linear mapping (left) and polynomial degree $3$ mapping (right) for a vertical section of the Sierra profile height in Section \ref{ssec:trex}.}
	\label{fig:mappings}
\end{figure}

We consider here a simpler approach based on the local polynomial interpolation of the curved geometry. We define as mappings the transformations from the reference element to elements in the physical space. Let \(\hat{\mathbf{x}}\) be a point in the reference element, namely the unit square \([0,1] \times [0,1] \subset \mathbb{R}^2.\) Hence, each mapping is a function \(\mathbf{F}_{K}\) such that \(\mathbf{x} = \mathbf{F}_{K}\left(\hat{\mathbf{x}}\right)\). We assume that the function \(\mathbf{F}_{K}\) is invertible. A relevant quantity is the Jacobian of the transformation \(\mathbf{J}_{K}\left(\hat{\mathbf{x}}\right) = \grad\mathbf{F}_{K}\left(\hat{\mathbf{x}}\right)\). Requiring that \(\mathbf{F}_{K}\) is invertible is equivalent to assuming \(\mathbf{J}_{K}\left(\hat{\mathbf{x}}\right)\) has nonzero determinant \cite{gurtin:1982}. For a generic function \(\phi(\mathbf{x})\), which can be real-valued, vector-valued or tensor-valued, the mapping acts therefore as follows:
\begin{equation}
	\phi\left(\mathbf{x}\right) = \phi\left(\mathbf{F}_{K}\left(\hat{\mathbf{x}}\right)\right)
	=\hat{\phi}\left(\hat{\mathbf{x}}\right).
\end{equation}
The DG method is characterized by integrals over elements and over faces that share two elements (see e.g. \cite{giraldo:2020} for a general presentation of the method). Using a simple change of variables, it is possible to express integrals over a cell \(K\) as integral over the reference element \(\hat{K}\). More specifically, the following relation holds \cite{arnold:2010, gurtin:1982}:
\begin{equation}\label{eq:integral_transform}
	\int_{K} \phi\left(\mathbf{x}\right)dx = \int_{\hat{K}} \hat{\phi}\left(\hat{\mathbf{x}}\right)\left|\text{det } \mathbf{J}_{K}\left(\hat{\mathbf{x}}\right)\right|d\hat{x},
\end{equation}  
where \(\text{det } \mathbf{J}_{K}\left(\hat{\mathbf{x}}\right)\) is the determinant of the Jacobian and \(d\hat{x}\) is the volume form in the reference element. Analogous considerations hold for face integrals. Note also that the transformation of differential forms, such as gradients of scalar functions, denoted by \(\mathbf{v}\), and gradients of vector fields, denoted by \(\mathbf{T}\), follows the general rule
\begin{equation}
	\mathbf{v}\left(\mathbf{x}\right) = \mathbf{A}\left(\hat{\mathbf{x}}\right)\hat{\mathbf{v}}\left(\hat{\mathbf{x}}\right) \qquad \mathbf{T}\left(\mathbf{x}\right) = \mathbf{A}\left(\hat{\mathbf{x}}\right)\hat{\mathbf{T}}\left(\hat{\mathbf{x}}\right)\mathbf{B}\left(\hat{\mathbf{x}}\right).
\end{equation}
The differential forms \(\mathbf{A}\) and \(\mathbf{B}\) are determined by the kind of object being transformed, as discussed in \cite{arnold:2010} (see also \cite{arnold:2015} for an analysis of finite element differential forms on curvilinear cubic meshes). For the sake of clarity, we report the transformation for the gradient of a shape function \(\varphi\), which reads as follows:
\begin{equation}
	\grad\varphi\left(\mathbf{x}\right) = \mathbf{J}_{K}^{-1}\left(\hat{\mathbf{x}}\right)\hat{\grad}\hat{\varphi}\left(\hat{\mathbf{x}}\right).
\end{equation} 
The integral in \eqref{eq:integral_transform} is then approximated by a quadrature rule, so as to obtain
\begin{equation}\label{eq:quadrature_formula}
	\int_{K} \phi\left(\mathbf{x}\right)dx \approx \sum_{q} \hat{\phi}\left(\hat{\mathbf{x}}_{q}\right)\left|\text{det }\mathbf{J}_{K}\left(\hat{\mathbf{x}}_{q}\right)\right|w_{q},
\end{equation}
where \(\hat{\mathbf{x}}_{q}\) are the nodes and \(w_{q}\) are the corresponding weights of the chosen quadrature formula. A visual impression of orography representation using high-order and low-order mapping is presented in Figure \ref{fig:mappings}. Note that the transformation depends on the location of the physical element, therefore it is different for each element.

%%%%%%%%%%%%%%%%%%%%%%%%%%%%% Results %%%%%%%%%%%%%%%%%%%%%%%%%%%%%%%%
\section{Numerical results}
\label{sec:num} \indent

The technique outlined in Section \ref{sec:high_order} is employed in a number of two-dimensional benchmarks concerning flows over orography. Several analytical profiles for the bottom boundary have been considered in literature. We focus here on two of them:
\begin{itemize}
	\item The \textit{versiera of Agnesi} \cite{bonaventura:2000, klemp:1978}
	\begin{equation}\label{eq:versiera_Agnesi}
		h(x) = \frac{h_{m}}{1 + \left(\frac{x - x_{c}}{a_{c}}\right)^2},
	\end{equation} 
	with \(h_{m}\) being the height of the orographic profile and \(a_{c}\) representing its half-width. 
	\item The five-peak mountain range profile \cite{schar:2002}
	\begin{equation}\label{eq:schar_profile}
		h(x) = h_{m}\exp\left[-\left(\frac{x}{a_{c}}\right)^{2}\right]\cos^{2}\left(\frac{\pi x}{\lambda_{c}}\right),
	\end{equation}
	where \(\lambda_{c}\) is a characteristic length scale.
\end{itemize}
Wall boundary conditions are employed for the bottom boundary, whereas non-reflecting boundary conditions are needed for the top boundary and for lateral boundaries. These are achieved by applying Rayleigh damping in the regions close to the boundary, with the following coefficient \cite{melvin:2019, orlando:2023a}:
\begin{equation}
	\lambda = 
	\begin{cases}
		0, \qquad & \text{if } z < z_{B} \\ \overline{\lambda}\sin^{2}\left[\frac{\pi}{2}\left(\frac{z - z_{B}}{z - z_{T}}\right)\right] \qquad & \text{if } z \ge z_{B}.
	\end{cases} 
\end{equation}
Here, \(z_{B}\) denotes the height at which the damping starts and \(z_{T}\) is the top height of the considered domain. Analogous definitions apply for the two lateral boundaries. The classical Gal-Chen height-based coordinates \cite{galchen:1975} are adopted to obtain a terrain-following mesh in Cartesian coordinates:
\begin{equation}\label{eq:galchen}
	z = \xi + \frac{z_{T} - \xi}{z_{T}}h(x),
\end{equation}
where \(\xi\) is the vertical coordinate of the rectangular domain before the transformation. Transformation \eqref{eq:galchen} leads to a terrain-following mesh at the bottom boundary, i.e. \(\xi = 0\), and to a horizontal top boundary, i.e. \(\xi = z_{T}\). 

A simple model for turbulent vertical diffusion for NWP applications, originally proposed in \cite{louis:1979} and also discussed e.g. in \cite{bonaventura:2014}, is employed in some configurations to achieve stable solutions. The nonlinear diffusivity \(\kappa\) has the form
\begin{equation}\label{eq:turbulent_diffusion}
	\kappa\left(\frac{\partial u}{\partial z}, \frac{\partial\theta}{\partial z}\right) = l^{2}\left|\frac{\partial u}{\partial z}\right|f\left(Ri\right),
\end{equation}
with \(l\) being a mixing length and \(Ri\) denoting the Richardson number given by
\begin{equation}
	Ri = \frac{g}{\theta_{0}}\frac{\frac{\partial\theta}{\partial z}}{\left|\frac{\partial u}{\partial z}\right|^{2}}.
\end{equation}
Here, \(\theta_{0}\) is a reference temperature, while the function \(f\left(Ri\right)\) is defined as
\begin{equation}
	f\left(Ri\right) = \left(1 + b\left|Ri\right|\right)^{\beta},
\end{equation}
where
\begin{equation}
	\begin{cases}
		\beta = -2, b = 5 \qquad &\text{if } Ri > 0 \\
		\beta = \frac{1}{2}, b = 20 \qquad &\text{if } Ri < 0.
	\end{cases}
\end{equation}
We also recall the definition of the vertical flux of horizontal momentum (henceforth ``momentum flux'') \cite{smith:1979}. This is an important diagnostic quantity, often used to verify if a correct orographic response is established in numerical experiments. It is defined as
\begin{equation}\label{eq:momentum_flux_isolated_orography}
	m(z) = \int_{-\infty}^{\infty} \overline{\rho}(z)u'(x,z)w'(x,z)dx.
\end{equation} 
Here, \(\overline{\rho}\) is the background density, whereas \(u'\) and \(w'\) denote the deviation from the background state of the horizontal and vertical velocity, respectively. Relation \eqref{eq:momentum_flux_isolated_orography} is valid only for an isolated orography profile with a constant background field \(\bar{u}\) \cite{smith:1979}. Hence, we also compute the full vertical flux of horizontal momentum, given by
\begin{equation}\label{eq:momentum_flux}
	m(z) = \int_{-\infty}^{\infty} \left(\bar{\rho}\left(z\right) + \rho'\left(x, z\right)\right)\left(\bar{u} + u'\left(x, z\right)\right)w'\left(x,z\right)dx,
\end{equation}
Table \ref{tab:parameters} reports the parameter values used in the test cases in this work. For time and space discretizations, we consider the IMEX-DG scheme proposed in \cite{orlando:2023b, orlando:2022} and successfully employed for atmospheric dynamics in \cite{orlando:2023a, orlando:2024}. We consider piecewise polynomials of degree \(r = 4\) for the finite element space, unless differently stated. The solver is implemented in the framework of the \texttt{deal.II} library \cite{arndt:2023, bangerth:2007}, which natively supports high-order polynomial mappings in addition to linear mappings.

\begin{table}[h!]
	\centering
	\footnotesize
	\begin{tabularx}{0.85\columnwidth}{lrrrXXr}
		\toprule
		\textbf{Test case} & $\boldsymbol{\Delta}\mathbf{t}$ & $\mathbf{T_{f}}$ & \textbf{Domain} & \textbf{Damping} & \textbf{Damping} & $\overline{\boldsymbol{\lambda}}\boldsymbol{\Delta}\mathbf{t}$ \\
		& $\boldsymbol{[}\SI{}{\second}\boldsymbol{]}$ & $\boldsymbol{[}\SI{}{\hour}\boldsymbol{]}$ & $\boldsymbol{[}\SI{}{\kilo\meter} \times \SI{}{\kilo\meter}\boldsymbol{]}$ & \textbf{layer} $\boldsymbol{(}x\boldsymbol{)}$ & \textbf{layer} $\boldsymbol{(}z\boldsymbol{)}$ \\
		\midrule
		LHMW & 2.5 & 15 & $240 \times 30$ & (0,80), (160,240) & (15,30) & 0.3 \\
		\midrule
		NLNHMW & 1 & 5 & $40 \times 20$ & (0,10), (30,40) & (9,20) & 0.15 \\ 
		\midrule
		Sch{\"a}r & 5 & 10 & $100 \times 30$ & (-50,-20), (20,50) & (20,30) & 1.2 \\
		\midrule
		NST & 0.5 & 6 & $100 \times 20$ & (0,20), (80,100) & (9,20) & 0.15 \\
		\midrule
		T-REX & 0.75 & 4 & $100 \times 26$ & (0,50), (350,400) & (20,26) & 0.15 \\
		\bottomrule
	\end{tabularx}
	\caption{Model parameters for the test cases, see main text for details. LHMW: linear hydrostatic mountain wave (Section \ref{ssec:hydrostatic}). NLNHMW: nonlinear non-hydrostatic mountain wave (Section \ref{ssec:non_hydrostatic}). Sch{\"a}r: Sch{\"a}r profile (Section \ref{ssec:schar}). NST: non-smooth orography (Section \ref{ssec:non_smooth}). T-REX: Sierra profile, T-REX experiment (Section \ref{ssec:trex}). The intervals where the damping layers are applied are in units of $\SI{}{\kilo\meter}$.}
	\label{tab:parameters}
\end{table}

%%%%%%%%%%%%%%%%%%%%%%%%%% Linear hydrostatic %%%%%%%%%%%%%%%%%%%%%%%
\subsection{Hydrostatic flow over a hill}
\label{ssec:hydrostatic} \indent

We first consider the linear hydrostatic configuration employed, e.g., in \cite{giraldo:2008, orlando:2023a}. The initial state consists of a constant mean flow with \(\bar{u} = \SI{20}{\meter\per\second}\) and of an isothermal background profile with temperature \(\bar{T} = \SI{250}{\kelvin}\). Finally, the initial profile of the Exner pressure is given by 
\begin{equation}
	\bar{\Pi} = \exp\left(-\frac{g}{c_{p}\bar{T}}z\right),
\end{equation}
with \(c_{p} = \frac{\gamma}{\gamma - 1}R\) denoting the specific heat at constant pressure. The bottom boundary is described by \eqref{eq:versiera_Agnesi}, with \(h_{m} = \SI{1}{\meter}, x_{c} = \SI{120}{\kilo\meter},\) and \(a_{c} = \SI{10}{\kilo\meter}\). 

For the construction of the boundary elements, we consider polynomial degrees 2 (i.e., a parabolic profile) and 4, and we compare the results with those obtained employing a linear mapping. The computational mesh is composed by \(100 \times 60\) elements, leading to a resolution of \(\SI{600}{\meter}\) along the horizontal direction and of \(\SI{125}{\meter}\) along the vertical direction. Note that, here and in the following, the effective resolution is computed dividing the size of the element along each direction by the polynomial degree, \(r = 4\) in this case. From linear theory \cite{smith:1979}, the analytical momentum flux is given by
\begin{equation}\label{eq:analytic_momentum_hydro}
	m^{H} = -\frac{\pi}{4}\overline{\rho}_{s}\overline{u}_{s} N h_{m}^{2},
\end{equation}
where \(\overline{\rho}_{s}\) and \(\overline{u}_{s}\) denote the surface background density and velocity, respectively, and \(N\) is the buoyancy frequency. 

\begin{figure}[h!]
	\centering
	\includegraphics[width=0.7\textwidth]{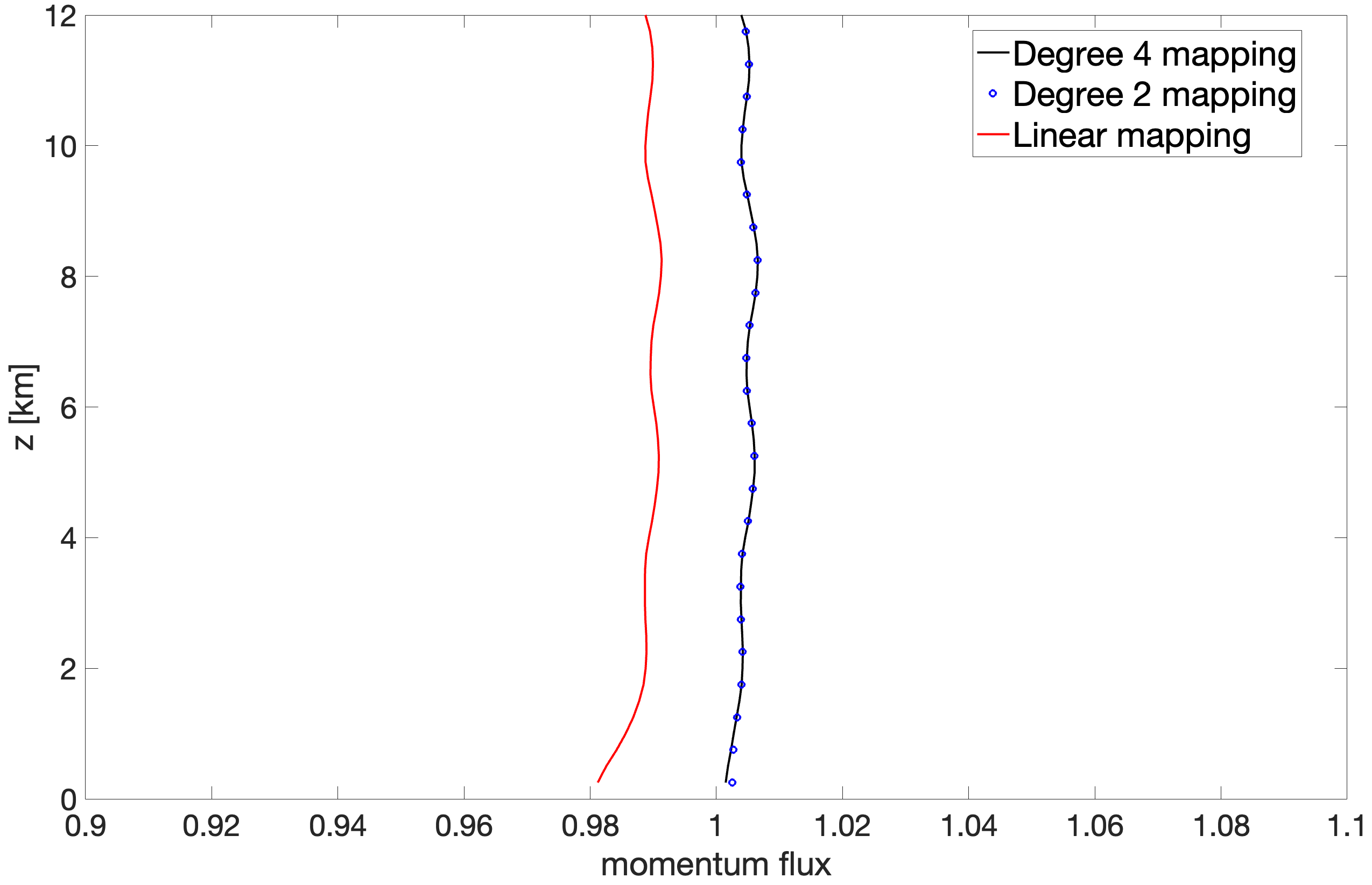}
	\caption{Linear hydrostatic flow over a hill, computed normalized momentum flux at $t = T_{f} = \SI{15}{\hour}$ using a mapping with polynomial degree $4$ (black line), a mapping with polynomial degree $2$ (blue dots), and the linear mapping (red line).}
	\label{fig:linear_hydro_momentum}
\end{figure}

\begin{figure}[h!]
	\centering
	\includegraphics[width=0.7\textwidth]{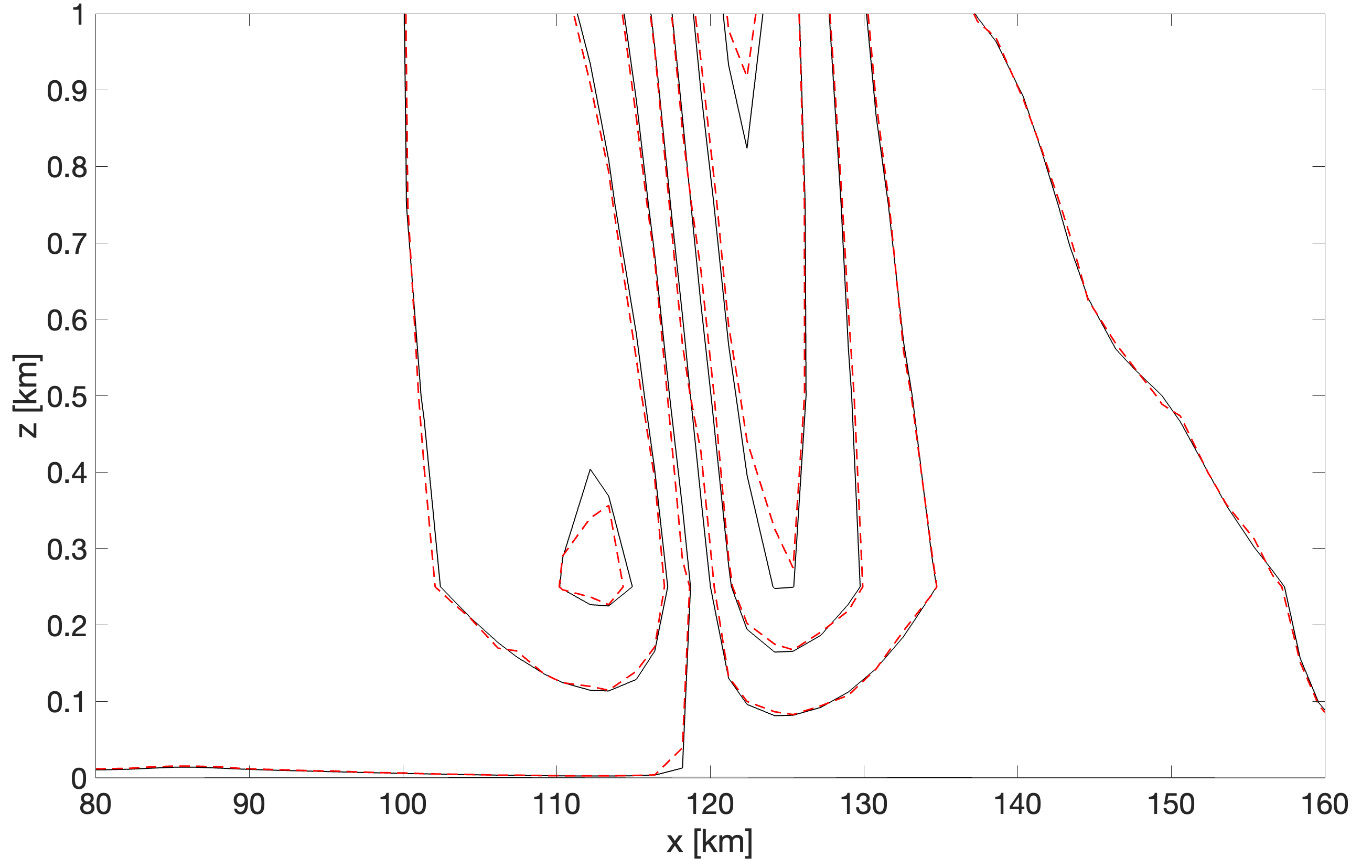}
	\caption{Linear hydrostatic flow over a hill, vertical velocity at $t = T_{f} = \SI{15}{\hour}$ using the high-order mapping with polynomial degree 4 (continuous black lines) and the linear mapping (red dashed lines). Contours in the range $\SI[parse-numbers=false]{[-4.0, 4.0] \times 10^{-3}}{\meter\per\second}$ with a $\SI[parse-numbers=false]{5 \times 10^{-4}}{\meter\per\second}$ contour interval.}
	\label{fig:linear_hydro_contours_comparison}
\end{figure}

The momentum flux at final time normalized by \(m^{H}\) and computed using curved elements for the bottom boundary provides an improved description of the orographic response compared to the same quantity computed using linear mapping (Figure \ref{fig:linear_hydro_momentum}). Moreover, increasing the polynomial degree of the mapping does not lead to further improvement in the accuracy of the results, meaning that the parabolic profile is already an excellent approximation for orographic profile \eqref{eq:versiera_Agnesi}. Similar considerations apply to the vertical velocity variable, albeit with slight differences (Figure \ref{fig:linear_hydro_contours_comparison}). The wall-clock time needed employing high-order mappings is essentially the same as that needed using the linear mapping (Table \ref{tab:overhead_linear_hydro}). We refer to Section \ref{ssec:WT_times} for further considerations on the computational cost of high-order mappings.

\begin{table}[h!]
	\centering
	\footnotesize
	\begin{tabularx}{0.4\columnwidth}{crc}
		\toprule
		Configuration & WT$[\SI{}{\second}]$ & Overhead \\
		\midrule
		Linear mapping & 10500 & - \\
		\midrule
		Degree 2 mapping & 10400 & -0.95\% \\
		\midrule
		Degree 4 mapping & 10600 & 0.95\% \\
		\bottomrule
	\end{tabularx}
	\caption{Linear hydrostatic flow over a hill: wall-clock times (WT) for the linear mapping and the high-order mappings simulations using a computational mesh composed by $N_{el} = 100 \times 60 = 6000$ elements. The overhead using the high-order mapping is computed with respect to the WT of the simulation employing the linear mapping.}
	\label{tab:overhead_linear_hydro}
\end{table}

%%%%%%%%%%%%%%%%%%%%%%% Non-hydrostatic flow %%%%%%%%%%%%%%%%%%%%%%%%%%
\subsection{Non-Hydrostatic flow over a hill}
\label{ssec:non_hydrostatic} \indent

Next, we consider the nonlinear non-hydrostatic configuration employed, e.g., in \cite{orlando:2023a, orlando:2024}. Here, the background velocity is \(\bar{u} = \SI{13.28}{\meter\per\second}\), whereas the initial state is described by the following potential temperature \(\bar{\theta}\) and Exner pressure \(\bar{\Pi}\):
\begin{equation}\label{eq:nonhydorstatic_init}
	\bar{\theta} = T_{ref}\exp\left(\frac{N^{2}}{g}z\right) \qquad \bar{\Pi} = 1 + \frac{g^{2}}{c_{p}T_{ref}N^{2}}\left[\exp\left(-\frac{N^{2}}{g}z\right) - 1\right],
\end{equation}
where \(T_{ref} = \SI{273}{\kelvin}\) denotes the surface temperature and \(N = \SI{0.02}{\per\second}\) denotes the constant Brunt-V{\"a}is{\"a}l{\"a} frequency. The bottom boundary is described again by \eqref{eq:versiera_Agnesi}, with \(h_{m} = \SI{450}{\meter}, x_{c} = \SI{20}{\kilo\meter},\) and \(a_{c} = \SI{1}{\kilo\meter}\). We consider again elements of polynomial degree equal to 2 and to 4, and we compare the results with those achieved with a linear mapping. The computational mesh is composed by \(50 \times 50\) elements, yielding a resolution of \(\SI{200}{\meter}\) along the horizontal direction and of \(\SI{100}{\meter}\) along the vertical direction. 

\begin{figure}[h!]
	\centering
	\includegraphics[width=0.7\textwidth]{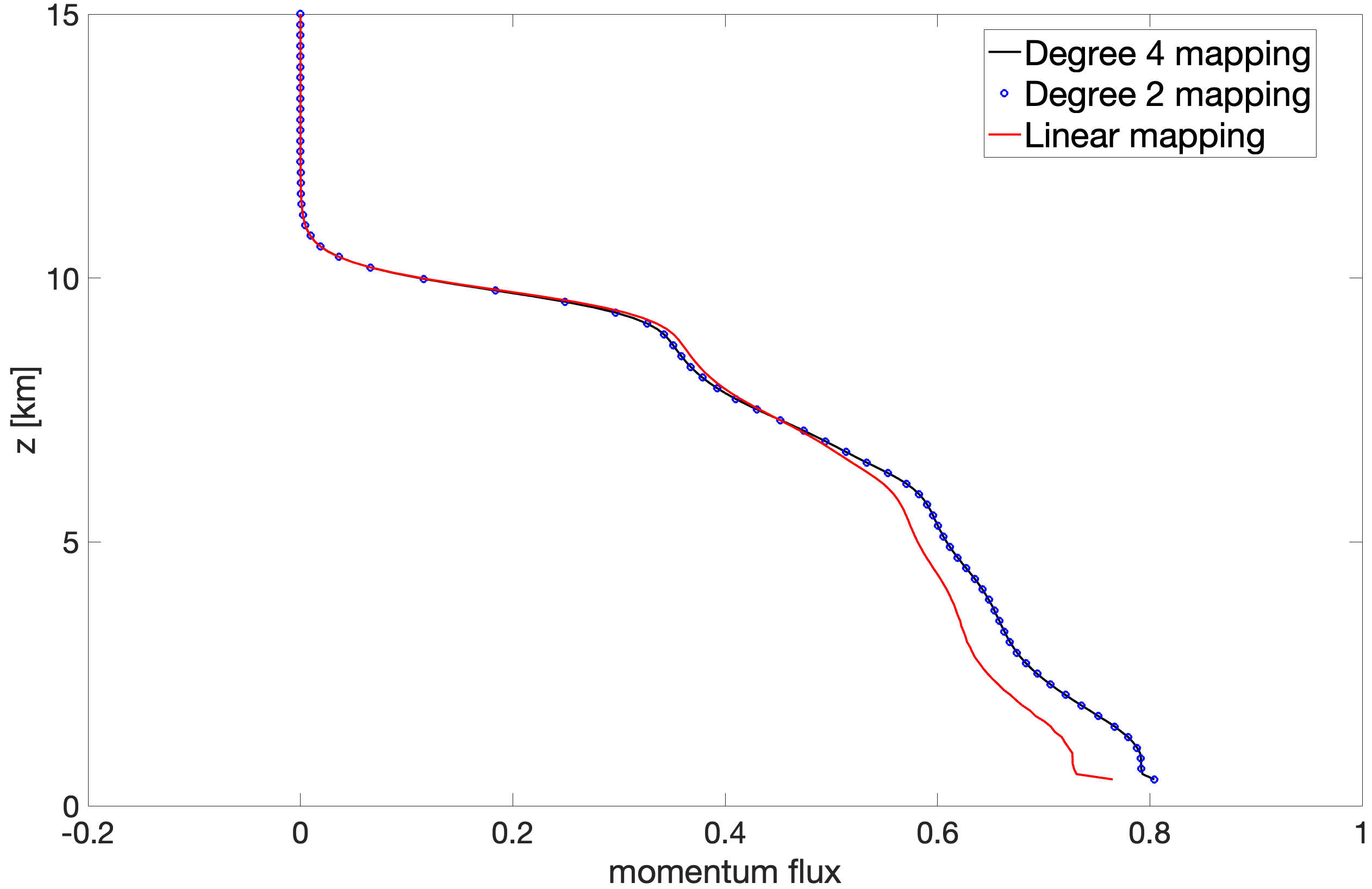}
	\caption{Nonlinear non-hydrostatic flow over a hill, comparison of normalized momentum flux at $t = T_{f} = \SI{5}{\hour}$ using a mapping with polynomial degree $4$ (black line), a mapping with polynomial degree $2$ (blue dots), and a linear mapping (red line).}
	\label{fig:nonlinear_nonhydro_momentum}
\end{figure}

The use of different polynomial mappings leads to different values of the normalized momentum flux at \(t = T_{f}\) in the lower part of the computational domain (Figure \ref{fig:nonlinear_nonhydro_momentum}). The maximum relative difference is around $8.2\%$, meaning that the geometric error associated with the description of the bottom boundary with a linear mapping is not negligible. Analogous considerations to those reported for the linear hydrostatic test case in Section \ref{ssec:hydrostatic} are valid for the vertical velocity (Figure \ref{fig:linear_hydro_contours_comparison}). Finally, for what concerns the computational cost, analogous considerations to those reported in Section \ref{ssec:hydrostatic} still hold. We refer again to Section \ref{ssec:WT_times} for further considerations on the computational cost.

\begin{table}[h!]
	\centering
	\footnotesize
	\begin{tabularx}{0.4\columnwidth}{crc}
		\toprule
		Configuration & WT$[\SI{}{\second}]$ & Overhead \\
		\midrule
		Linear mapping & 3320 & - \\
		\midrule
		Degree 2 mapping & 3600 & 8.4\% \\
		\midrule
		Degree 4 mapping & 3480 & 4.8\% \\
		\bottomrule
	\end{tabularx}
	\caption{Nonlinear non-hydrostatic flow over a hill: wall-clock times (WT) for the linear mapping and the high-order mappings simulations using a computational mesh composed by $N_{el} = 50 \times 50 = 2500$ elements. The overhead using the high-order mapping is computed with respect to the WT of the simulation employing the linear mapping.}
	\label{tab:overhead_nonlinear_nonhydro}
\end{table}

\begin{figure}[h!]
	\centering
	\includegraphics[width=0.7\textwidth]{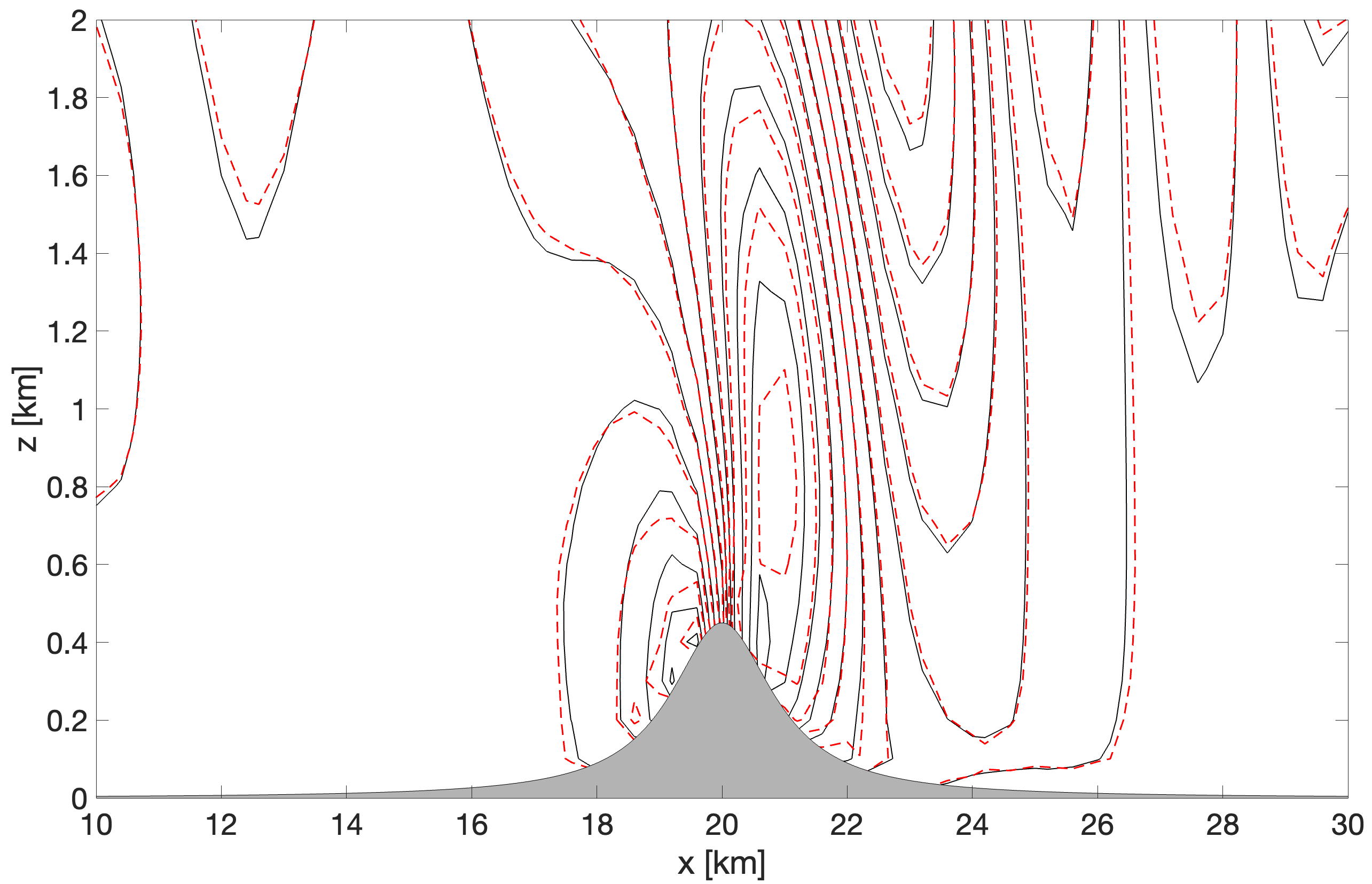}
	\caption{Nonlinear non-hydrostatic flow over a hill, vertical velocity at $t = T_{f} = \SI{5}{\hour}$ using the high-order mapping with polynomial degree 4 (continuous black lines) and the linear mapping (red dashed lines). Contours in the range $\SI[parse-numbers=false]{[-4.2, 4.0]}{\meter\per\second}$ with a $\SI[parse-numbers=false]{0.586}{\meter\per\second}$ contour interval.}
	\label{fig:nonlinear_nonhydro_contours_comparison}
\end{figure}

%%%%%%%%%%%%%%%%%%%%%%% Schar profile %%%%%%%%%%%%%%%%%%%%%%%%%%%%%%%%
\subsection{Flow over Sch{\"a}r profile}
\label{ssec:schar} \indent

\begin{figure}[h!]
	\centering
	\begin{subfigure}{0.475\textwidth}
		\centering
		\includegraphics[width = 0.95\textwidth]{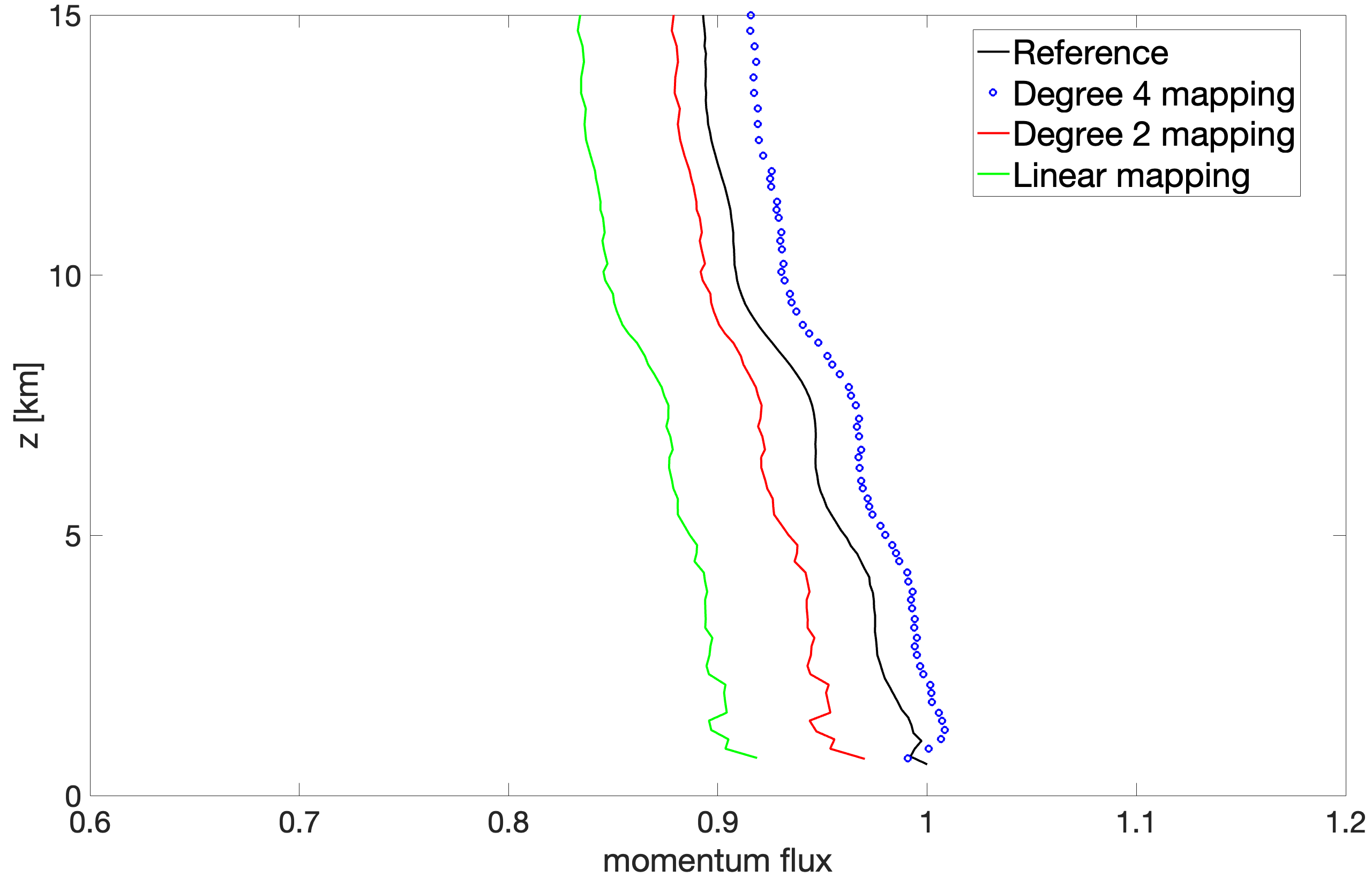}
	\end{subfigure}
	\begin{subfigure}{0.475\textwidth}
		\centering
		\includegraphics[width = 0.95\textwidth]{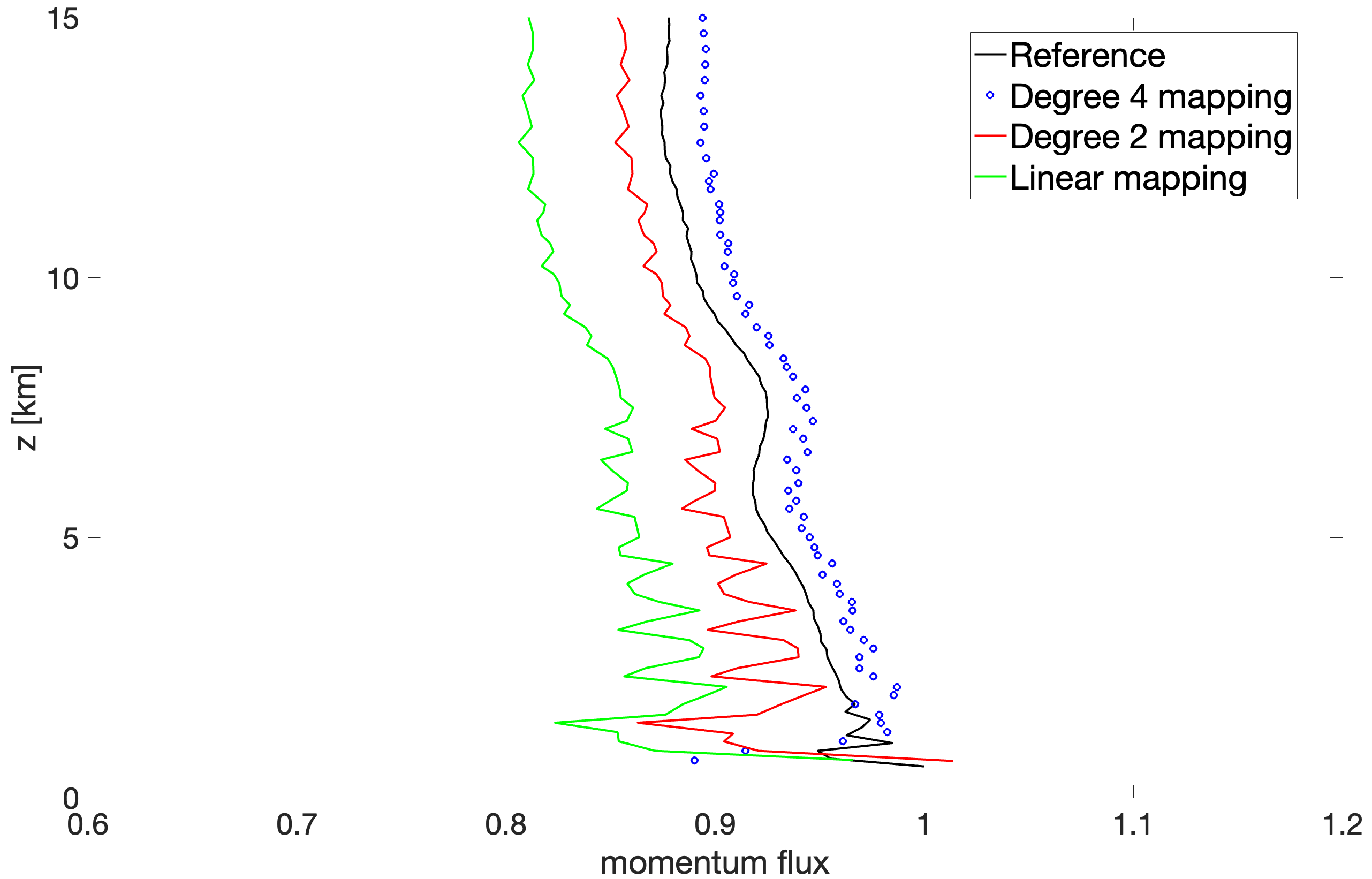}
	\end{subfigure}
	\caption{Flow over Sch{\"a}r profile, normalized momentum flux computed with formula \eqref{eq:momentum_flux_isolated_orography} (left) and formula \eqref{eq:momentum_flux} (right) at $t = T_{f} = \SI{10}{\hour}$, using polynomial degree $4$ mapping (blue line), polynomial degree $2$ mapping (red line), linear mapping (green line). The black line denotes a reference solution with a $200 \times 50$ mesh using a linear map (black line). In all cases, the momentum flux is normalized using the value at $z = \SI{600}{\meter}$ obtained with the reference solution.} 
	\label{fig:schar_momentum}
\end{figure}

Next, we consider the more complex five-peak idealized mountain range \eqref{eq:schar_profile} originally proposed in \cite{schar:2002} (see also \cite{melvin:2019, orlando:2023a}). We choose \(h_{m} = \SI{250}{\meter}, a_{c} = \SI{5}{\kilo\meter}\), and \(\lambda_{c} = \SI{4}{\kilo\meter}\). The initial state has the expression \eqref{eq:nonhydorstatic_init}, with surface temperature \(T_{ref} = \SI{288}{\kelvin}\), constant buoyancy frequency \(N = \SI{0.01}{\per\second}\), and a background horizontal velocity \(\bar{u} = \SI{10}{\meter\per\second}\). The domain is \(\Omega = \SI[parse-numbers=false]{\left(-50, 50\right) \times \left(0, 30\right)}{\kilo\meter}\). The mesh is composed by \(50 \times 25\) elements, leading to a resolution of \(\SI{500}{\meter}\) along the horizontal direction and of \(\SI{300}{\meter}\) along the vertical direction. A reference solution is computed using a mesh composed by \(200 \times 50\) elements and a linear mapping to describe the orography. We compute the momentum flux \eqref{eq:momentum_flux_isolated_orography} at \(t = T_{f}\) and we normalize it by the value obtained with the reference solution at \(z = \SI{600}{\meter}\) (Figure \ref{fig:schar_momentum}). It is important to remark that \eqref{eq:momentum_flux_isolated_orography} is valid only for an isolated orography with a constant background field \(\bar{u}\) \cite{smith:1979}. Hence, following \cite{guerra:2018}, we also compute the full vertical flux of horizontal momentum, given by
\begin{equation}
	m(z) = \int_{x_{min}}^{x_{max}} \left(\bar{\rho}\left(z\right) + \rho'\left(x, z\right)\right)\left(\bar{u} + u'\left(x, z\right)\right)w'\left(x,z\right)dx,
\end{equation}
with \(x_{min} = \SI{0}{\kilo\meter}\) and \(x_{max} = \SI{30}{\kilo\meter}\). 

\begin{figure}[h!]
	\centering
	\includegraphics[width=0.7\textwidth]{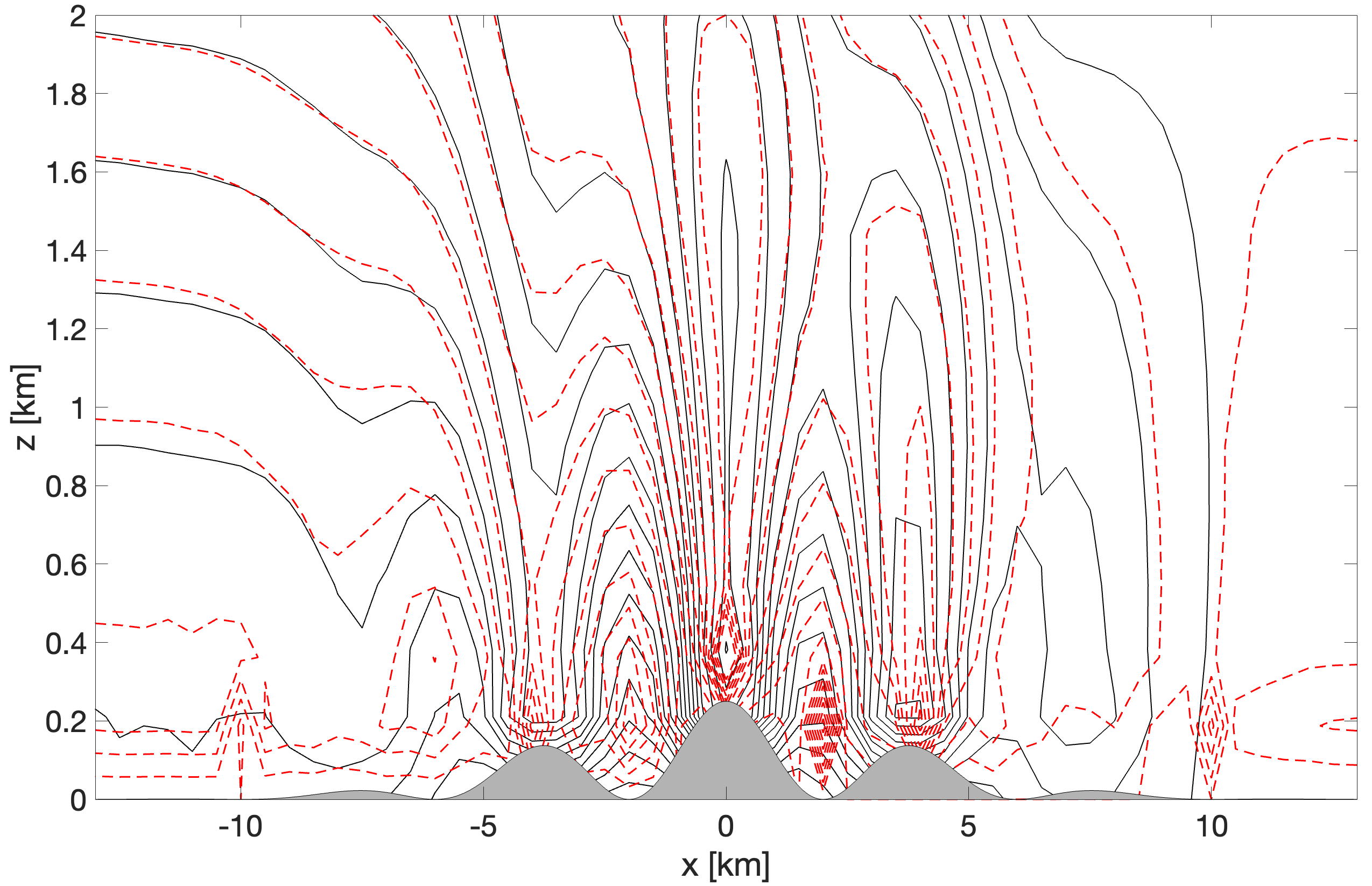}
	\caption{Flow over Sch{\"a}r profile, horizontal velocity deviation at $t = T_{f} = \SI{10}{\hour}$ using a mapping with polynomial degree $4$ (solid black contours) and a linear mapping (dashed red contours). Contours in the range $\SI[parse-numbers=false]{[-2, 2]}{\meter\per\second}$ with a $\SI[parse-numbers=false]{0.2}{\meter\per\second}$ contour interval.}
	\label{fig:schar_contours_comparison}
\end{figure}

Significant differences arise in the transfer of the momentum flux along the vertical direction (Figure \ref{fig:schar_momentum}). Moreover, for this test case, increasing the polynomial degree of the mapping yields significantly different results. This is due to the fact that \eqref{eq:schar_profile} is described by a non-polynomial function. Therefore, the accuracy in the orography description improves as the polynomial degree of the mapping increases. We also notice that the use of a polynomial degree equal to \(4\) reduces the oscillations in the profile of the momentum flux. Finally, high-order mappings avoid small oscillations close to the orography for the horizontal velocity deviation (Figure \ref{fig:schar_contours_comparison}) and this further emphasizes the importance of the use of high-order mappings. Analogous considerations to those reported in Section \ref{ssec:hydrostatic} hold for the computational time (Table \ref{tab:overhead_schar}).

\begin{table}[h!]
	\centering
	\footnotesize
	\begin{tabularx}{0.4\columnwidth}{crc}
		\toprule
		Configuration & WT$[\SI{}{\second}]$ & Overhead \\
		\midrule
		Linear mapping & 2030 & - \\
		\midrule
		Degree 2 mapping & 2140 & 5.4\% \\
		\midrule
		Degree 4 mapping & 1930 & -4.9\% \\
		\bottomrule
	\end{tabularx}
	\caption{Flow over Sch{\"a}r profile: wall-clock times (WT) for the linear mapping simulation and the high-order mappings simulations using a computational mesh composed by $N_{el} = 50 \times 25 = 1250$ elements. The overhead using the high-order mappings is computed with respect to the WT of the simulation employing the linear mapping.}
	\label{tab:overhead_schar}
\end{table}

%%%%%%%%%%%%%%%%%%%%%%%%%% Non-Smooth %%%%%%%%%%%%%%%%%%%%%%%%%%%%%%
\clearpage

\subsection{Non-smooth orography}
\label{ssec:non_smooth} \indent

In this Section, we modify one of the previous classical benchmarks by adding a non-smooth perturbation. More specifically, we consider the \textit{versiera of Agnesi} \eqref{eq:versiera_Agnesi}, with \(h_{m} = \SI{0.45}{\kilo\meter}, x_{c} = \SI{50}{\kilo\meter},\) and \(a_{c} = \SI{8}{\kilo\meter}\), and we add the perturbation
\begin{equation}
	h'\left(x\right) = 1 - 4\left|x - \left\lfloor x + \frac{1}{2}\right\rfloor\right|, 
\end{equation}
where \(\lfloor x\rfloor\) denotes the largest integer smaller or equal to \(x\) (floor function) and the length is in units of kilometers. Hence, the resulting orography reads as follows (Figure \ref{fig:non_smooth_profile_delta0,025}):
\begin{equation}\label{eq:non_smooth_orography}
	h(x) =
	\begin{cases}
		\frac{h_{m}}{1 + \left(\frac{x - x_{c}}{a_{c}}\right)^{2}} +  h_{m}\delta\left(1 - 4\left|x - \left\lfloor x + \frac{1}{2}\right\rfloor\right|\right) \qquad \text{if } \left|x - x_{c}\right| \le 2a_{c} \\
		\frac{h_{m}}{1 + \left(\frac{x - x_{c}}{a_{c}}\right)^{2}} \qquad \text{otherwise},
	\end{cases}
\end{equation}
with \(\delta = 0.025\). The perturbation has zero mean value, so that, if an increasingly strong filter is applied, the original smooth profile is recovered. Notice that 
\begin{equation}
	\frac{N a_{c}}{\bar{u}} \approx 12 \gg 1,
\end{equation}
which implies  that we are considering a hydrostatic regime.

\begin{figure}[h!]
	\centering
	\includegraphics[width = 0.7\textwidth]{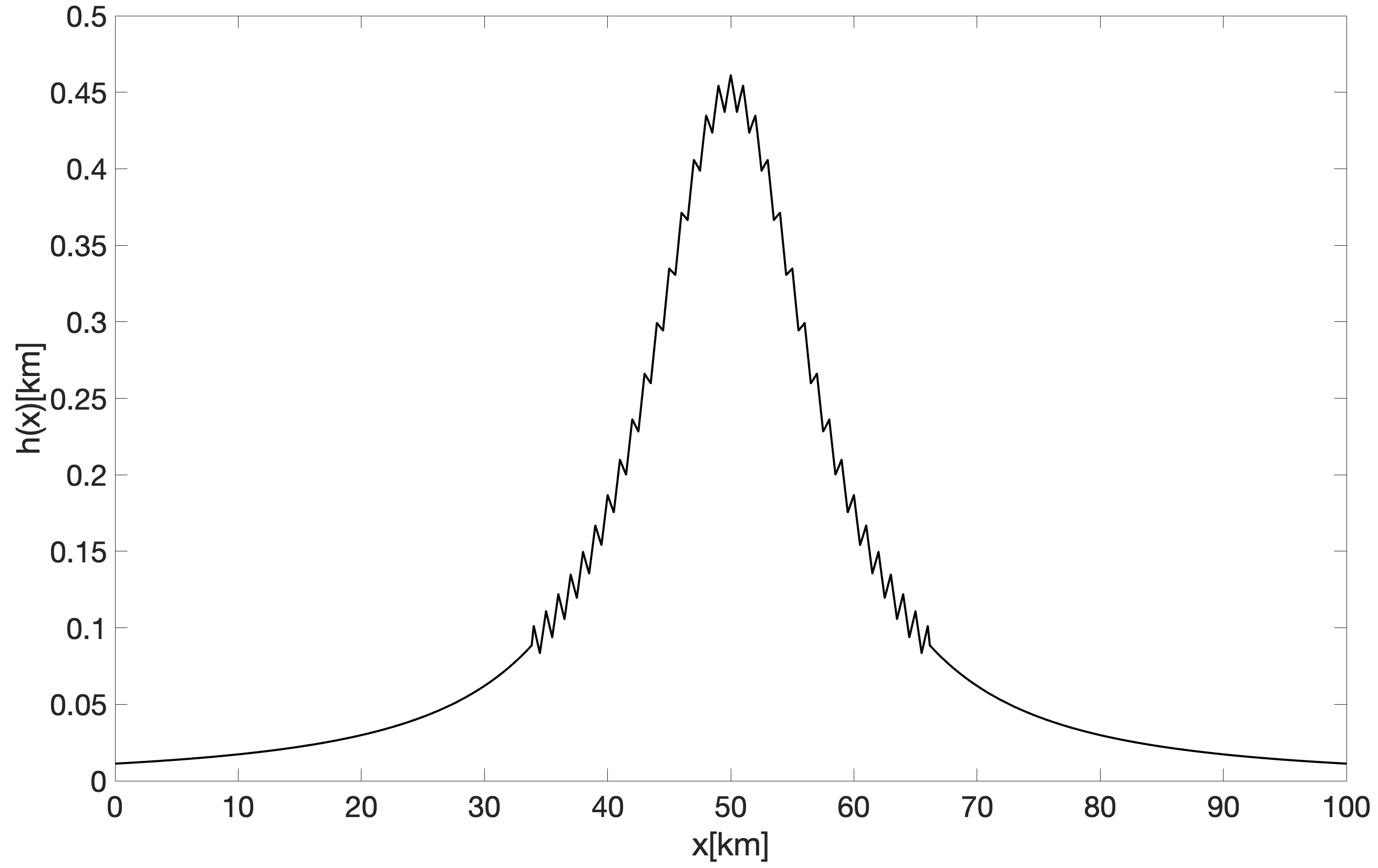}
	\caption{Non-smooth orography profile obtained with formula \eqref{eq:non_smooth_orography} using $\delta = 0.025$.}
	\label{fig:non_smooth_profile_delta0,025}
\end{figure}

The computational mesh is composed by \(100 \times 50\) elements, yielding a resolution of \(\SI{250}{\meter}\) along the horizontal direction and of \(\SI{100}{\meter}\) along the vertical direction. A reference solution is computed using a computational mesh composed by \(300 \times 50\) elements with a linear mapping, so as to allow a correct description of the non-smooth perturbation. The value obtained at \(z = \SI{550}{\meter}\) for the reference solution is adopted to normalize the momentum flux. Following the discussion in \cite{smith:1979}, the momentum flux \eqref{eq:momentum_flux_isolated_orography} is associated with a pressure difference which results in a net drag force on the mountain. For an isolated orography with a constant background field, one can show that \eqref{eq:momentum_flux_isolated_orography} is equivalent to
\begin{equation}
	m(z) = \int_{-\infty}^{\infty}p'\left(x,z\right)\frac{dh(x)}{dx}dx,
\end{equation}
with \(p'\) denoting the pressure perturbation with respect to the background state. Hence, the momentum flux is strongly influenced by the orography, and therefore it can be employed as an indicator to analyze the quality in the description of the orographic profile. 

The use of high-order mappings leads to results in good agreement with the reference solution, while a visible discrepancy arises with the linear mapping (Figure \ref{fig:non_smooth_normalized_momentum_flux_comparison_delta0,025}). Quantitatively, we compute the \(l^{2}\) relative error in the region \(\left[z_{1}, z_{2}\right]\), with \(z_{1} = \SI{1}{\kilo\meter}\) and \(z_{2} = \SI{9}{\kilo\meter}\), excluding the damping layer (Table \ref{tab:errors_non_smooth_orography_delta0,025}). The results on the momentum flux computed with \eqref{eq:momentum_flux_isolated_orography} show that the use of high-order mappings provides a better description of some small-scale features of the orography even at lower spatial resolution and leads to improved results in the development of lee waves and large-scale features. Analogous considerations hold for the vertical flux of horizontal momentum computed with \eqref{eq:momentum_flux}.
The overhead of the simulation employing the high-order mapping with respect to that using the linear mapping amounts to around $4.8\%$ in terms of wall-clock time (Table \ref{tab:errors_non_smooth_orography_delta0,025}).

\begin{table}[h!]
	\centering
	\footnotesize
	\begin{tabularx}{0.92\columnwidth}{cXXcXXc}
		\toprule
		\multirow{2}{*}{$t$} & \multicolumn{2}{c}{Linear mapping} & & \multicolumn{2}{c}{High-order mapping} \\
		\cmidrule(l){2-3}\cmidrule(l){5-7}
		& error $m(z)$ \eqref{eq:momentum_flux_isolated_orography} & error $m(z)$ \eqref{eq:momentum_flux} & & error $m(z)$ \eqref{eq:momentum_flux_isolated_orography} & error $m(z)$ \eqref{eq:momentum_flux} & Overhead \\
		\midrule
		$\frac{T_{f}}{2} = \SI{3}{\hour}$ & $6.37 \times 10^{-2}$ & $6.43 \times 10^{-2}$ & & $1.88 \times 10^{-2}$ & $1.90 \times 10^{-2}$ & \\
		\midrule
		$T_{f} = \SI{6}{\hour}$ & $3.48 \times 10^{-2}$ & $3.64 \times 10^{-2}$ & & $6.86 \times 10^{-3}$ & $7.46 \times 10^{-3}$ & 4.8\% \\
		\bottomrule
	\end{tabularx}
	\caption{Flow over non-smooth orography with $\delta = 0.025$ in \eqref{eq:non_smooth_orography}, $l^{2}$ relative errors on the normalized momentum flux computed with both \eqref{eq:momentum_flux_isolated_orography} and \eqref{eq:momentum_flux} using linear mapping and degree 3 (high-order) mapping. The relative error is computed with respect to the reference solution in the region $\left[z_{1}, z_{2}\right]$, with $z_{1} = \SI{1}{\kilo\meter}$ and $z_{2} = \SI{9}{\kilo\meter}$. The overhead using the high-order mapping is computed with respect to the WT of the simulation employing the linear mapping.}
	\label{tab:errors_non_smooth_orography_delta0,025}
\end{table}

\begin{figure}[h!]
	\centering
	\begin{subfigure}{0.475\textwidth}
		\centering
		\includegraphics[width = 0.95\textwidth]{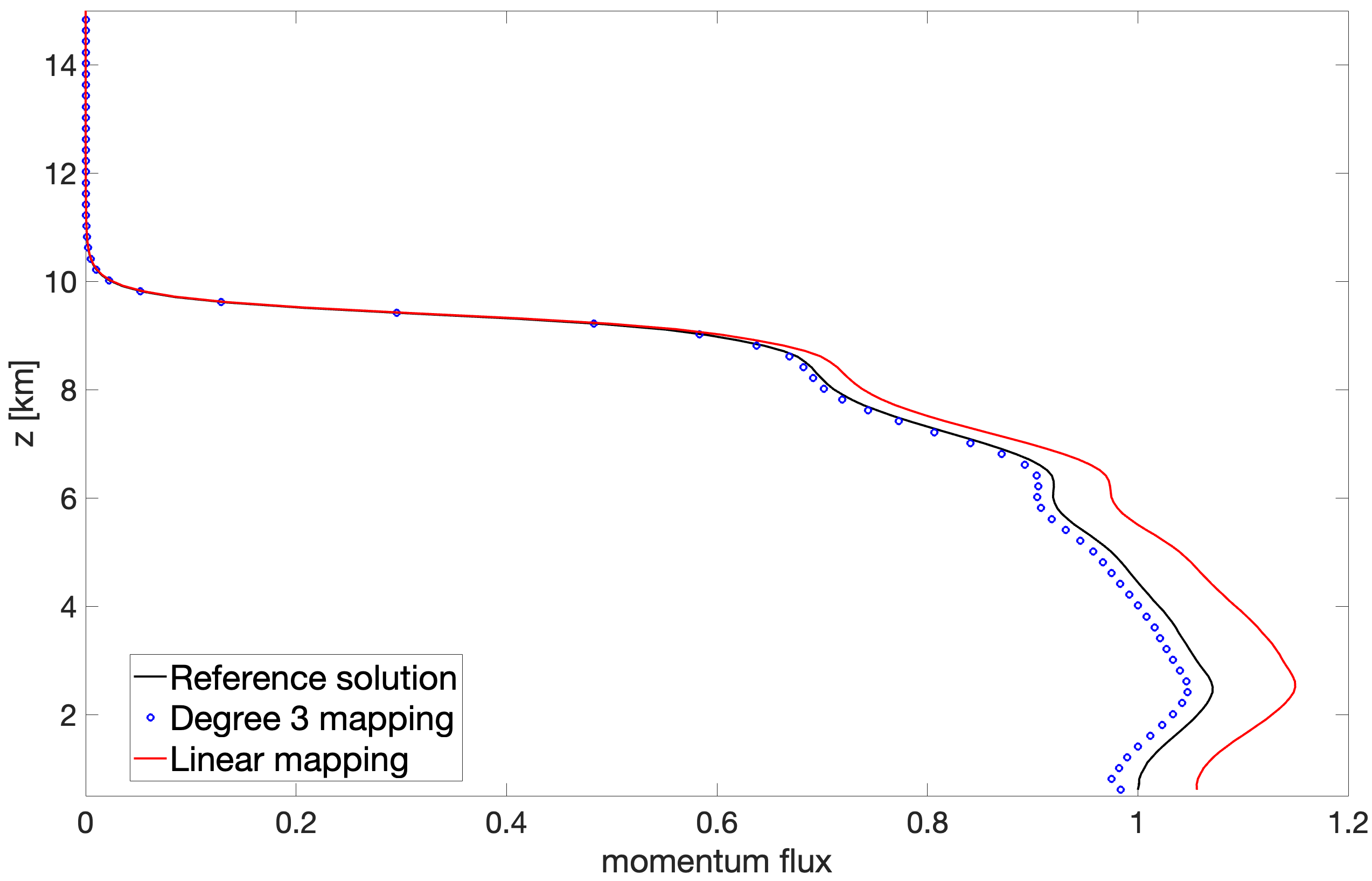}
	\end{subfigure}
	\begin{subfigure}{0.475\textwidth}
		\centering
		\includegraphics[width = 0.95\textwidth]{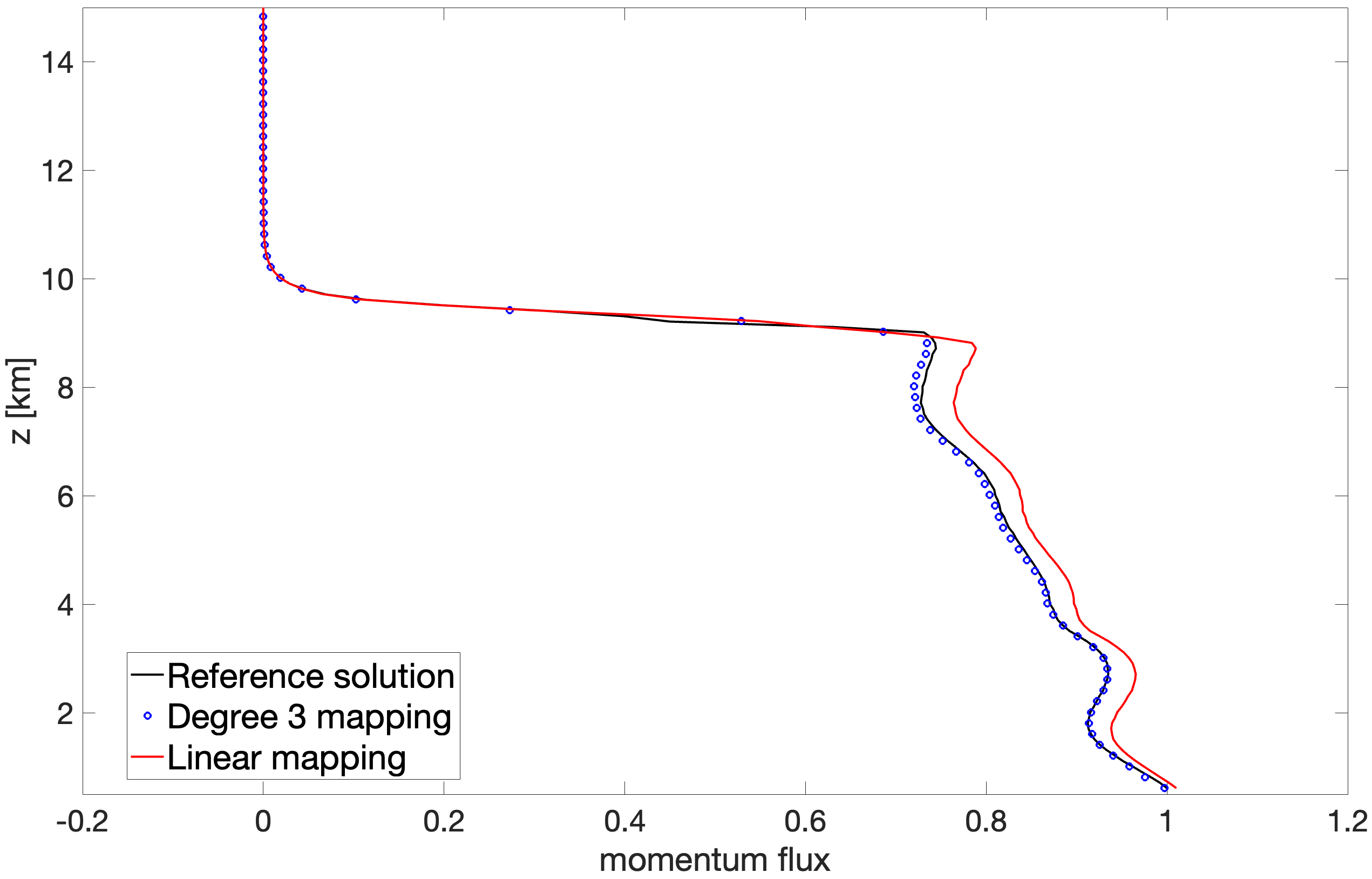}
	\end{subfigure}
	\caption{Flow over non-smooth orography with $\delta = 0.025$ in \eqref{eq:non_smooth_orography}, normalized momentum flux computed with \eqref{eq:momentum_flux_isolated_orography} at times $t = \frac{T_{f}}{2} = \SI{3}{\hour}$ (left panel) and $t = T_{f} = \SI{6}{\hour}$ (right panel). Results with polynomial degree 3 mapping (blue dots), linear mapping (red line), and reference results with a $300 \times 50$ elements mesh and linear mapping (black line). The momentum flux is normalized by the value obtained with the $300 \times 50$ elements mesh at $z = \SI{550}{\meter}$.}
	\label{fig:non_smooth_normalized_momentum_flux_comparison_delta0,025}
\end{figure}

Next, we increase the perturbation and we set \(\delta = 0.15\) in \eqref{eq:non_smooth_orography} (Figure \ref{fig:non_smooth_profile_delta0,15}). We also employ the simple model for turbulent vertical diffusion \eqref{eq:turbulent_diffusion} and we take \(l = \SI{50}{\meter}\) and \(\theta_{0} = \SI{273}{\kelvin}\) in \eqref{eq:turbulent_diffusion}.

\begin{figure}[h!]
	\centering
	\includegraphics[width = 0.7\textwidth]{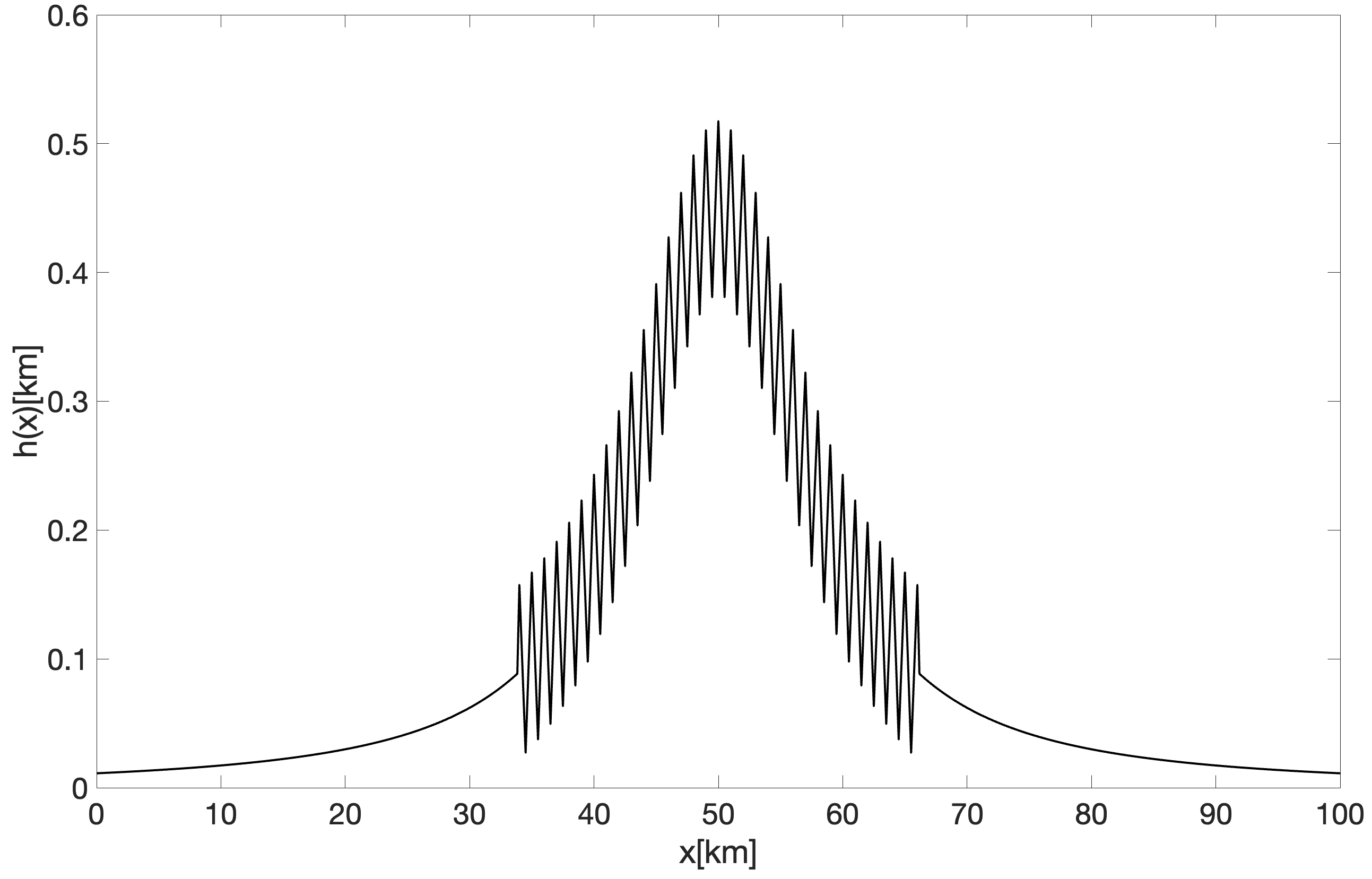}
	\caption{Non-smooth orography profile obtained with formula \eqref{eq:non_smooth_orography} using $\delta = 0.15$ in \eqref{eq:non_smooth_orography}.}
	\label{fig:non_smooth_profile_delta0,15}
\end{figure}

A comparison of the contours of the horizontal velocity deviation and of the potential temperature shows that high-order mapping results more closely match the reference results compared to linear mapping results (Figure \ref{fig:non_smooth_contours_delta0,15}). This is confirmed by the values of the normalized momentum flux at $t = \frac{T_{f}}{{2}}$ and at $t = T_{f}$ (Figure \ref{fig:non_smooth_normalized_momentum_flux_comparison_delta0,15} and Table \ref{tab:errors_non_smooth_orography_delta0,15}). In terms of relative \(l^{2}\) error with respect to the reference solution, the high-order mapping outperforms the linear mapping by an order of magnitude. Slightly larger values of the error of the momentum flux using the high-order mapping are obtained for larger values of the final time, due to the severity of the artificial orography profile (not shown). Moreover, the overhead of the simulation employing the high-order mapping with respect to that using the linear mapping amounts to $\approx 71\%$ in terms of wall-clock time (Table \ref{tab:errors_non_smooth_orography_delta0,15}). The possible reasons for this overhead will be discussed in Section \ref{ssec:WT_times}.

While a linear mapping might seem the most appropriate choice for the non-smooth, non-differentiable orography profile \eqref{eq:non_smooth_orography}, the use of a linear mapping on coarse meshes yields a poor description of small-scale orographic patterns. It is well known \cite{fritts:2022, kanehama:2019} that small-scale orographic patterns can have a significant impact on the development of lee waves and, more in general, on mountain wave-driven atmospheric processes. Improved representation of the orography significantly contributes to improvements in forecast skill at increasing resolutions.

\begin{table}[h!]
	\centering
	\footnotesize
	\begin{tabularx}{0.92\columnwidth}{cXXcXXc}
		\toprule
		\multirow{2}{*}{$t$} & \multicolumn{2}{c}{Linear mapping} & & & \multicolumn{2}{c}{High-order mapping} \\
		\cmidrule(l){2-3}\cmidrule(l){5-7}
		& error $m(z)$ \eqref{eq:momentum_flux_isolated_orography} & error $m(z)$ \eqref{eq:momentum_flux} & & error $m(z)$ \eqref{eq:momentum_flux_isolated_orography} & error $m(z)$ \eqref{eq:momentum_flux} & Overhead \\
		\midrule
		$\frac{T_{f}}{2} = \SI{3}{\hour}$ & $1.30$ & $1.31$ & & $1.72 \times 10^{-1}$ & $1.69 \times 10^{-1}$ & \\
		\midrule
		$T_{f} = \SI[parse-numbers=false]{6}{\hour}$ & $3.79 \times 10^{-1}$ & $3.76 \times 10^{-1}$ & & $3.26 \times 10^{-2}$ & $3.19 \times 10^{-2}$ & $71\%$ \\
		\bottomrule
	\end{tabularx}
	\caption{Flow over non-smooth orography with $\delta = 0.15$ in \eqref{eq:non_smooth_orography}, $l^{2}$ relative errors on the normalized momentum flux computed with both \eqref{eq:momentum_flux_isolated_orography} and \eqref{eq:momentum_flux}. The relative error is computed with respect to the reference solution in the region $\left[z_{1}, z_{2}\right]$, with $z_{1} = \SI{1}{\kilo\meter}$ and $z_{2} = \SI{9}{\kilo\meter}$. The overhead is computed with respect to the WT of the simulation employing the linear mapping.}
	\label{tab:errors_non_smooth_orography_delta0,15}
\end{table}

\begin{figure}[h!]
	\centering
	\begin{subfigure}{0.95\textwidth}
		\centering
		\includegraphics[width = 0.7\textwidth]{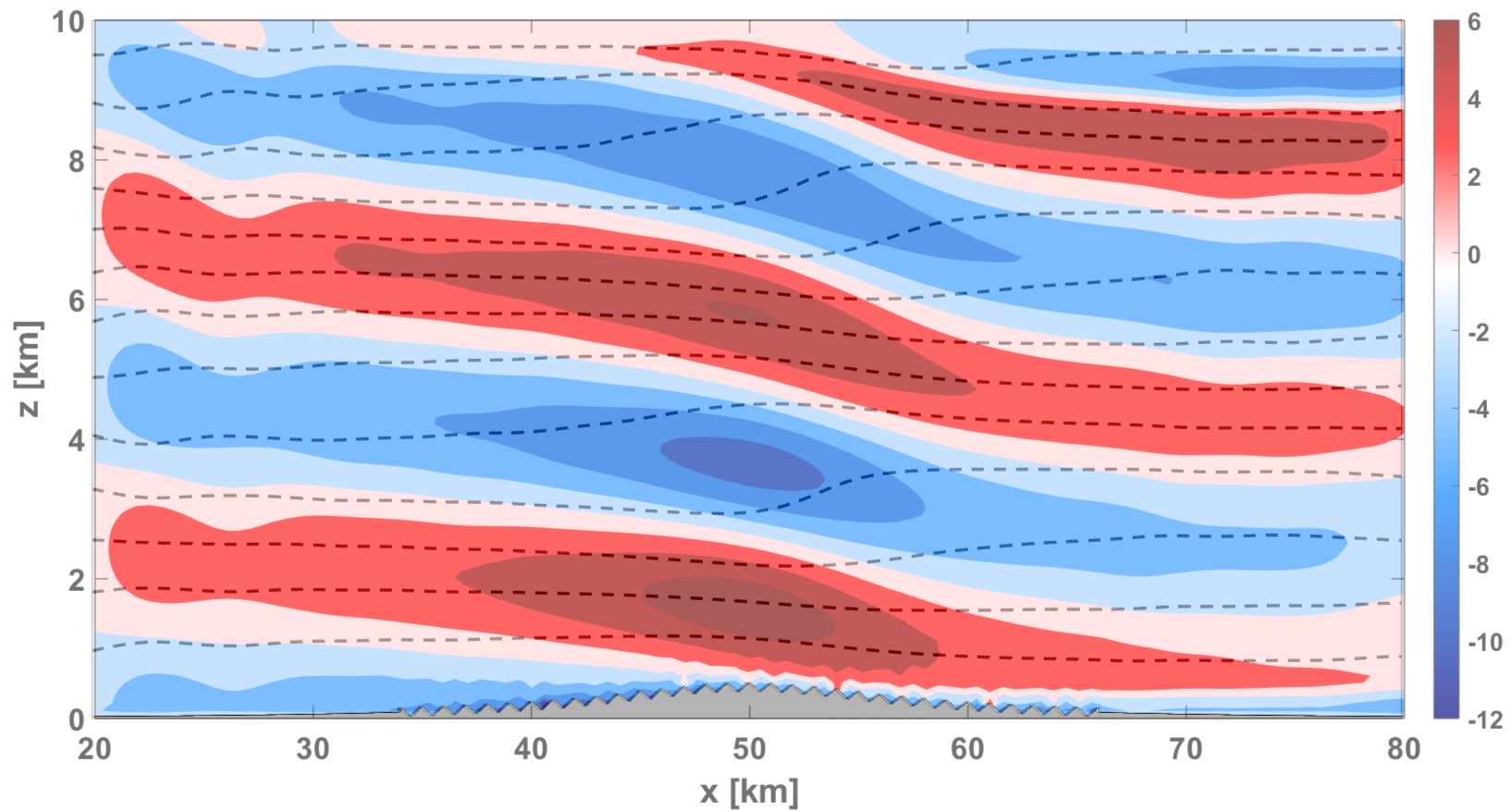}
	\end{subfigure}
	\begin{subfigure}{0.95\textwidth}
		\centering
		\includegraphics[width = 0.7\textwidth]{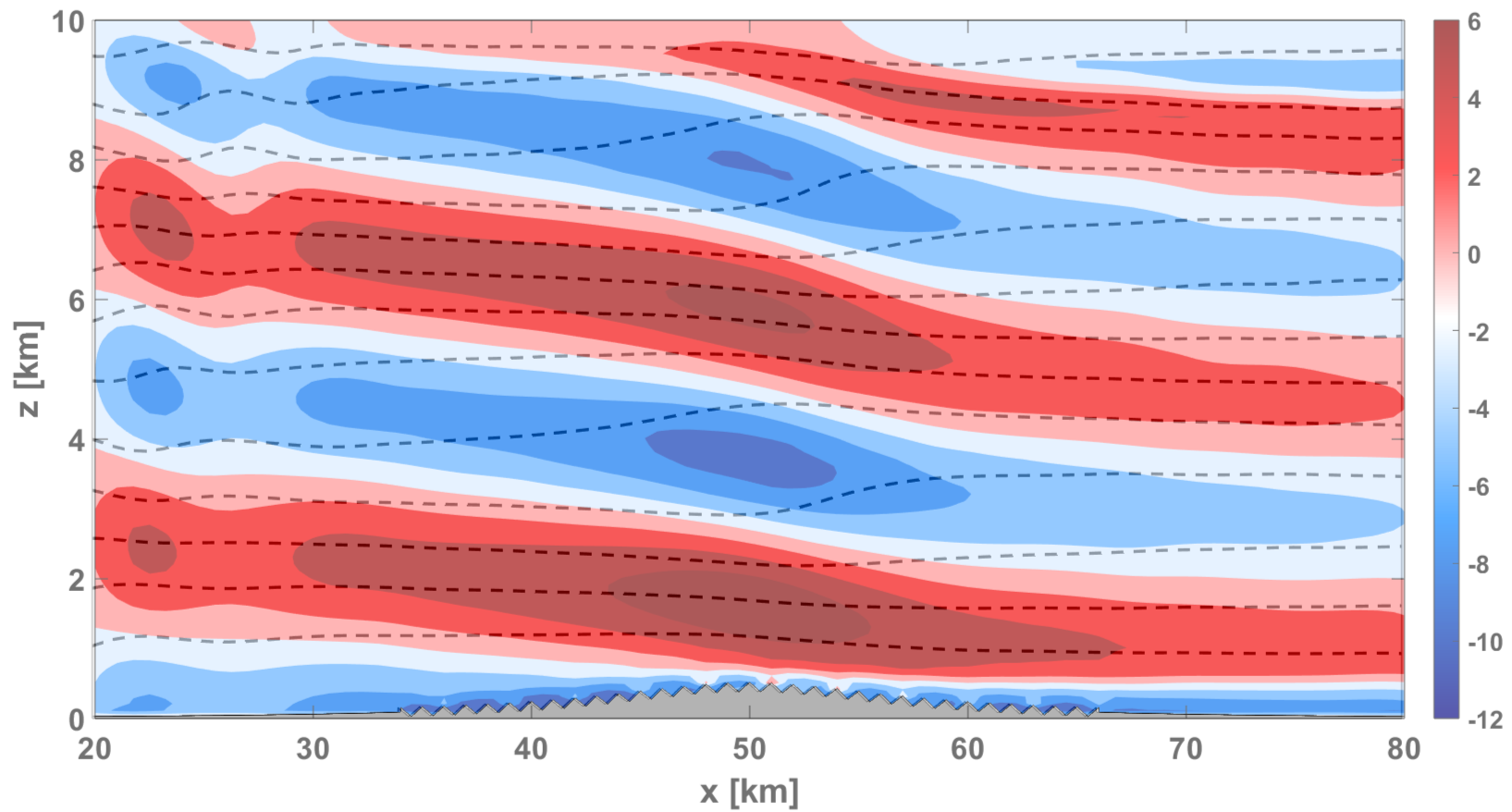}
	\end{subfigure}
	\begin{subfigure}{0.95\textwidth}
		\centering
		\includegraphics[width = 0.7\textwidth]{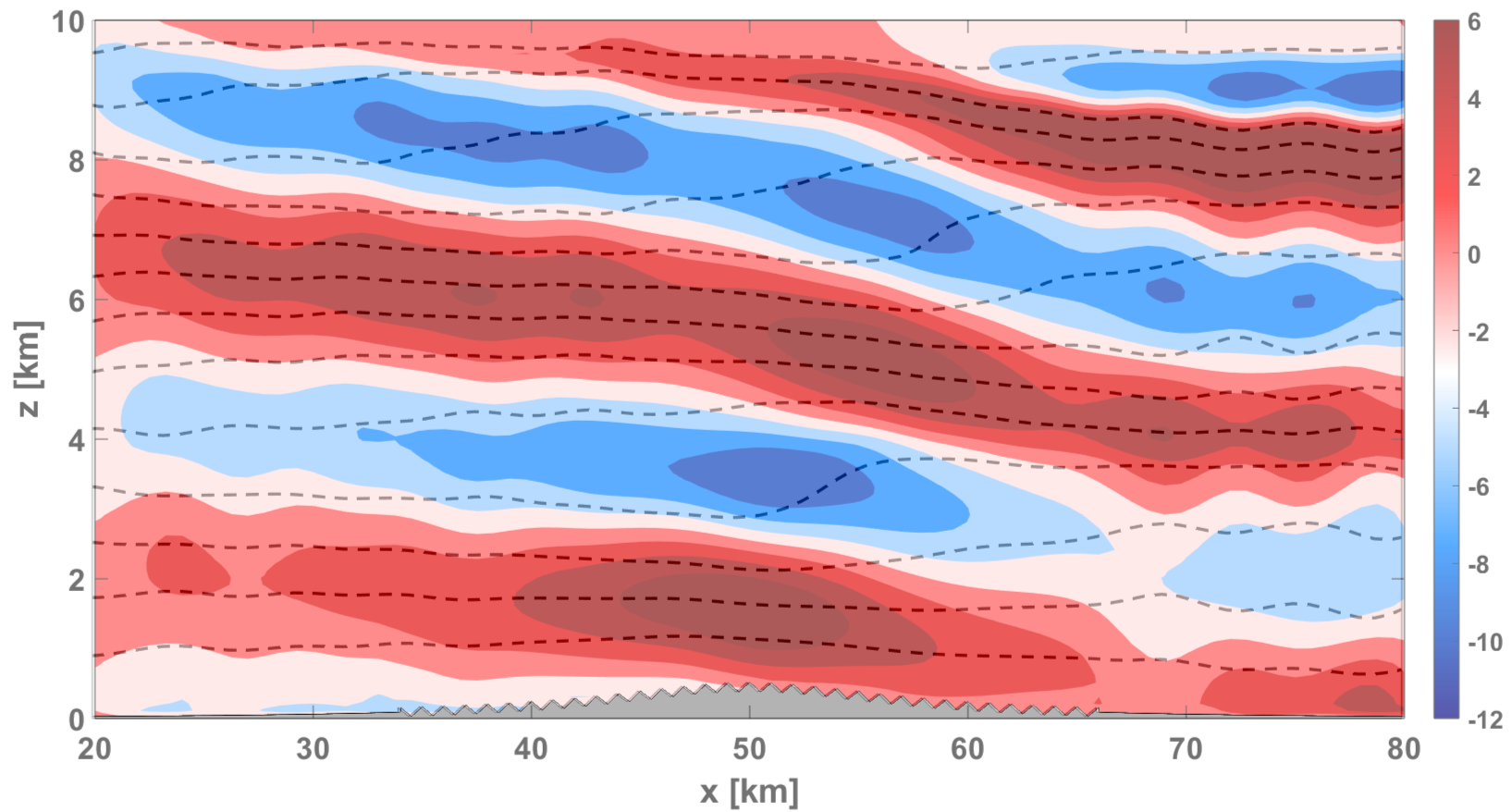}
	\end{subfigure}
	\caption{Flow over non-smooth orography with $\delta = 0.15$ in \eqref{eq:non_smooth_orography} at $t = T_{f} = \SI{6}{\hour}$. Top: Reference solution. Middle: Polynomial degree $3$ mapping. Bottom: Linear mapping. Horizontal velocity perturbation (colors), contours in the range $\SI[parse-numbers=false]{[-15,15]}{\meter\per\second}$ with a $\SI{2.5}{\meter\per\second}$ interval. Potential temperature (dashed lines), contours in the range $\SI[parse-numbers=false]{[273, 403]}{\kelvin}$ with a $\SI{10}{\kelvin}$ interval.}
	\label{fig:non_smooth_contours_delta0,15}
\end{figure}

\begin{figure}[h!]
	\centering
	\begin{subfigure}{0.475\textwidth}
		\centering
		\includegraphics[width = 0.95\textwidth]{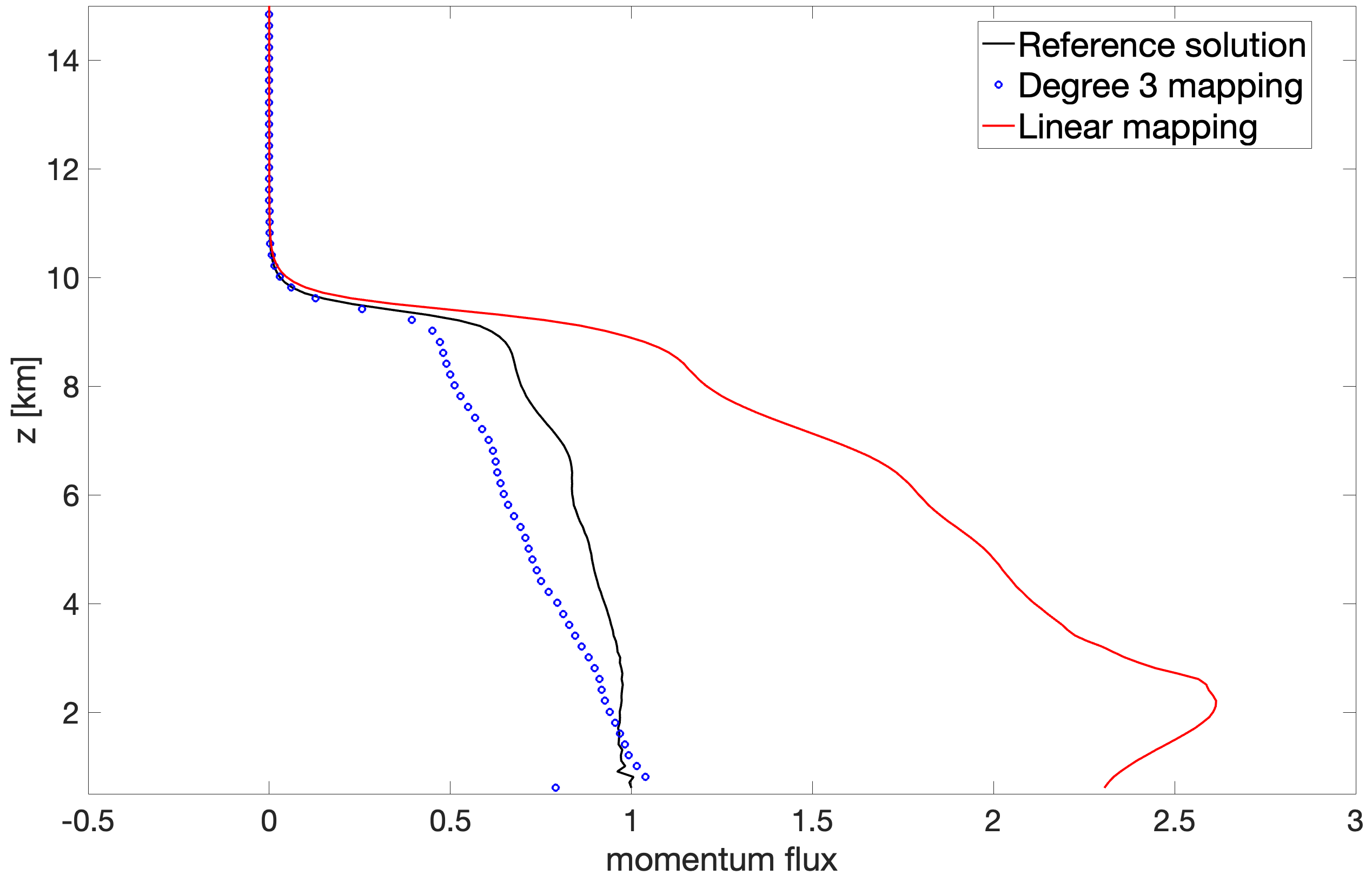}
	\end{subfigure}
	\begin{subfigure}{0.475\textwidth}
		\centering
		\includegraphics[width = 0.95\textwidth]{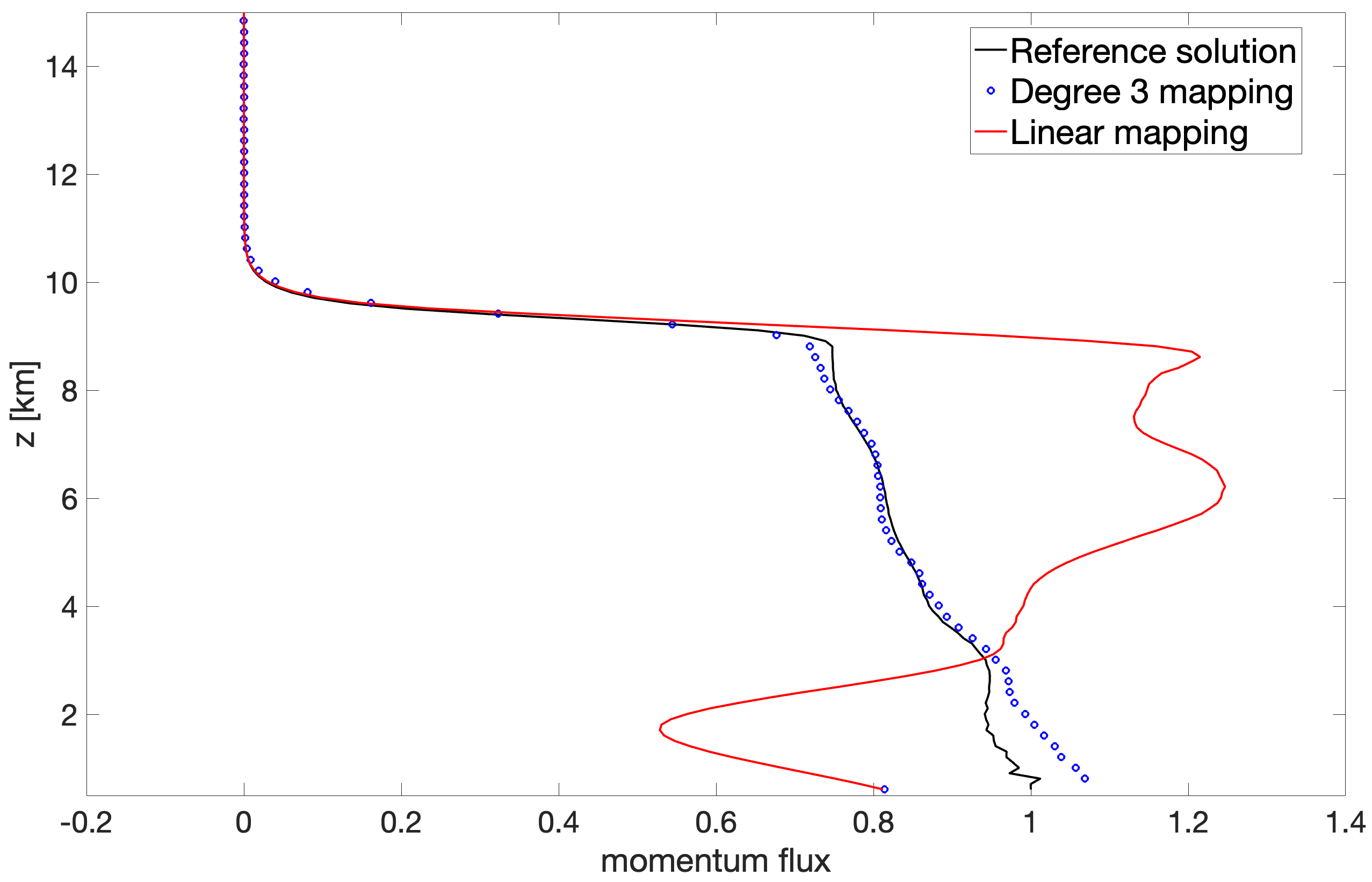}
	\end{subfigure}
	\caption{Flow over non-smooth orography with $\delta = 0.15$ in \eqref{eq:non_smooth_orography}, normalized momentum flux computed with \eqref{eq:momentum_flux_isolated_orography} at $t = \frac{T_{f}}{2}$ (left panel), $t = T_{f}$ (right panel). Results with polynomial degree 3 mapping (blue dots), linear mapping (red line), and reference results with a $300 \times 50$ elements mesh and linear mapping (black line). The momentum flux is normalized by the value at $z = \SI{550}{\meter}$ obtained with the $300 \times 50$ elements mesh.}
	\label{fig:non_smooth_normalized_momentum_flux_comparison_delta0,15}
\end{figure}

\subsubsection{Filtered non-smooth orography}

Next, we consider the results obtained smoothing the orography using filters. This approach seeks to avoid the resolution of small-scale features at a resolution close to that of the mesh, so as to avoid spurious oscillations and problems with physical parametrizations, see, e.g., the discussion in \cite{webster:2003}. Subgrid-scale orographic (gravity wave) drag parametrizations are employed in NWP and climate models to compensate the insufficient resolution of orographic features \cite{miller:1989, palmer:1986}. However, the interplay between resolved and parameterized orographic effects is critical, and improved results in simulations of mountain atmospheric processes can be obtained by increasing the resolution, see \cite{fritts:2022, kanehama:2019}, and the recent contribution of the authors on the use of non-conforming meshes \cite{orlando:2024}. 

In order to smooth the orography, we consider a discrete set of \(N\) values \(h_{i}\) computed from \eqref{eq:non_smooth_orography} and we apply two different filtering techniques: the moving average filter and the Raymond filter described in \cite{webster:2003}. We take \(N = 601\) equispaced points. For the moving average, we define the new discrete orography as
\begin{equation}\label{eq:ma}
	\hat{h}_{i} = \frac{1}{M+1}\sum_{i - \frac{M}{2}}^{i + \frac{M}{2}}h_i,
\end{equation}
with \(M + 1\) being the number of values employed for the average. We consider here \(M = 4,6\). The Raymond filter is instead a spectral filter and it is defined by specifying the Fourier coefficients of its response function as
\begin{equation}\label{eq:sf}
	\tilde{F}(\hat{\kappa}) = \frac{1}{1 + \varepsilon\tan^{m}\left(\frac{1}{2}\hat{\kappa}\Delta\right)},
\end{equation}
where \(\hat{\kappa}\) denotes the wave number, \(\varepsilon\) a suitable constant, and \(m\) a suitable exponent. The variable \(\Delta\) is related to the resolution in the description of the orography. Since the orography changes after the application of the filter, a reference solution using the linear mapping on the \(300 \times 50\) elements mesh is computed for each filtered configuration. A cubic spline interpolation \cite{deboor:1978} is employed to evaluate the orography at the points of this finer mesh.

\begin{figure}[h!]
	\centering
	\begin{subfigure}{0.9\textwidth}
		\centering
		\includegraphics[width=0.7\textwidth]{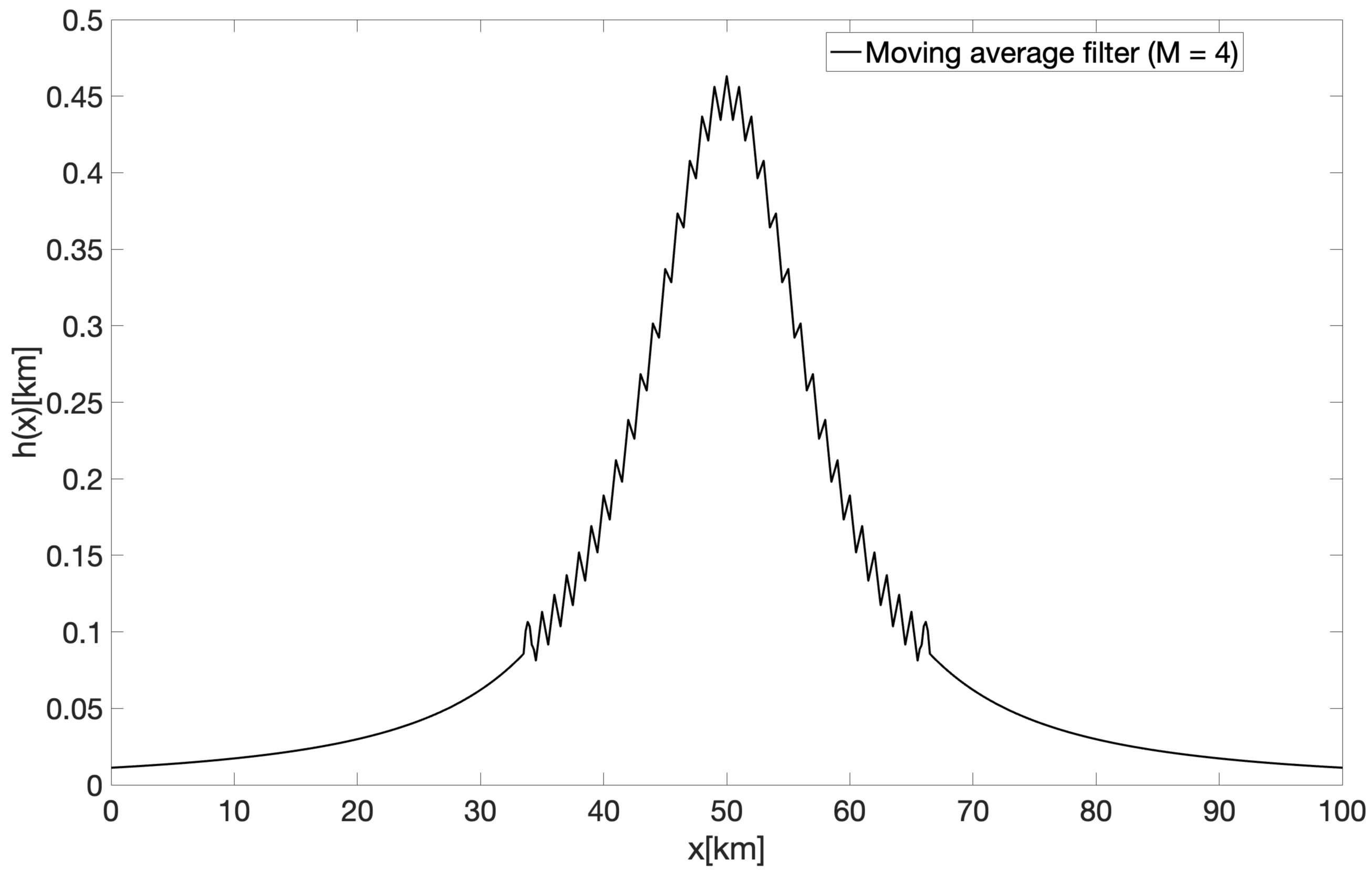}
	\end{subfigure}
	\begin{subfigure}{0.9\textwidth}
		\centering
		\includegraphics[width=0.7\textwidth]{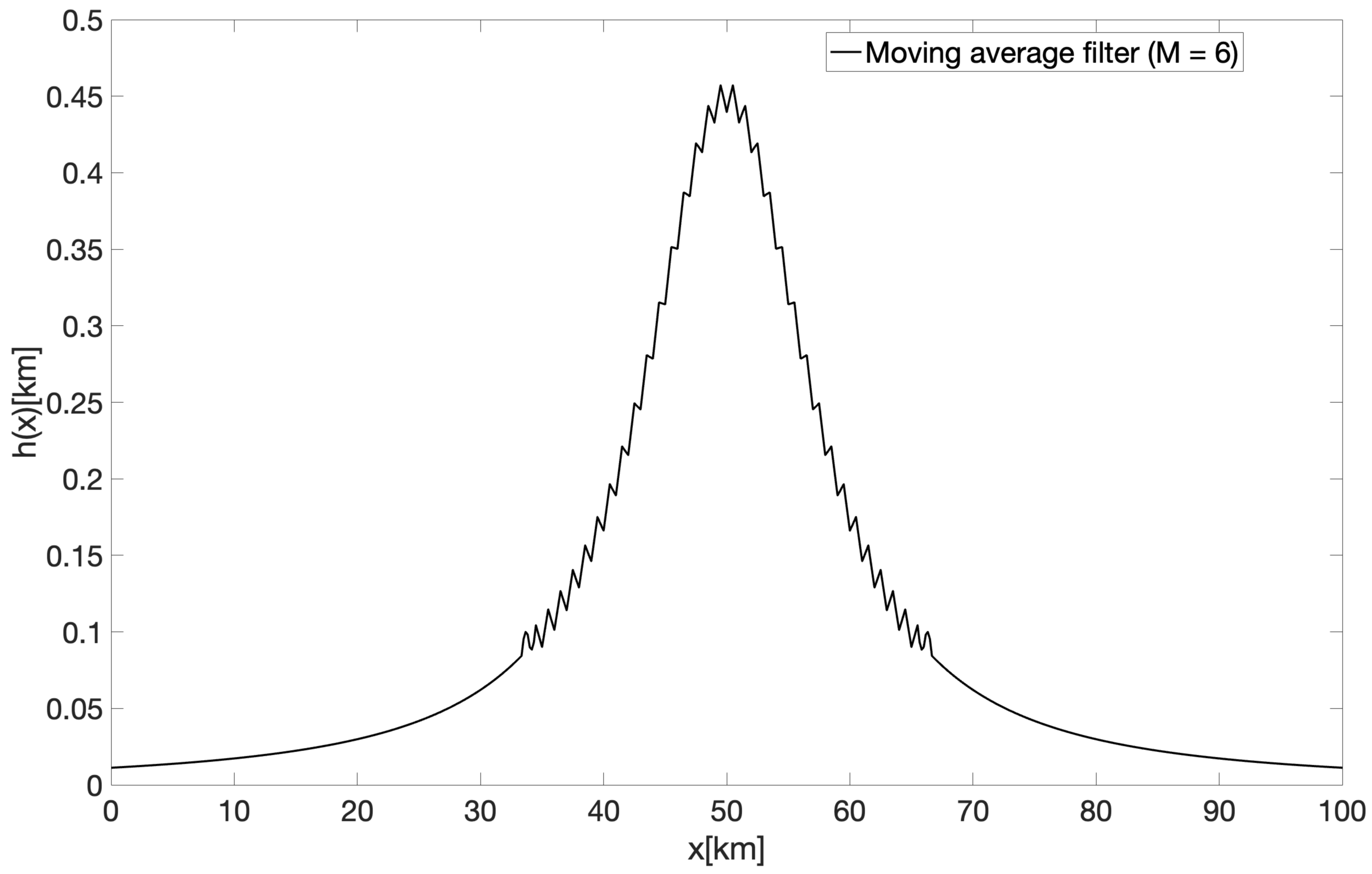}
	\end{subfigure}
	\caption{Non-smooth orography with $\delta = 0.15$ in formula \eqref{eq:non_smooth_orography}, smoothed profiles with moving-average filter \eqref{eq:ma}. Top: $M=4$. Bottom: $M=6$.}
	\label{fig:non_smooth_filtered_orographies_ma}
\end{figure}

\begin{figure}[h!]
	\centering
	\begin{subfigure}{0.475\textwidth}
		\centering
		\includegraphics[width = 0.95\textwidth]{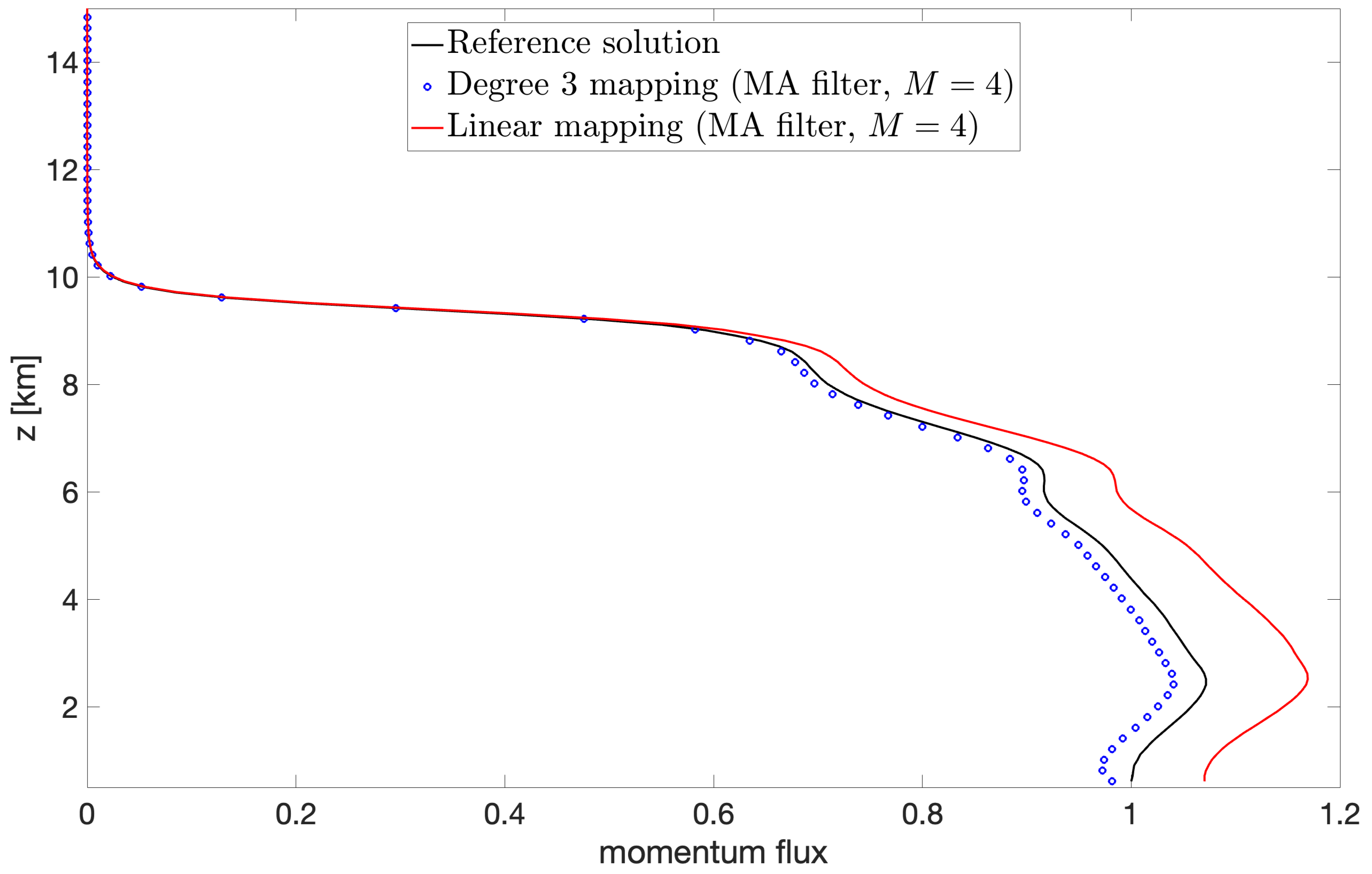}
	\end{subfigure}
	\begin{subfigure}{0.475\textwidth}
		\centering
		\includegraphics[width = 0.95\textwidth]{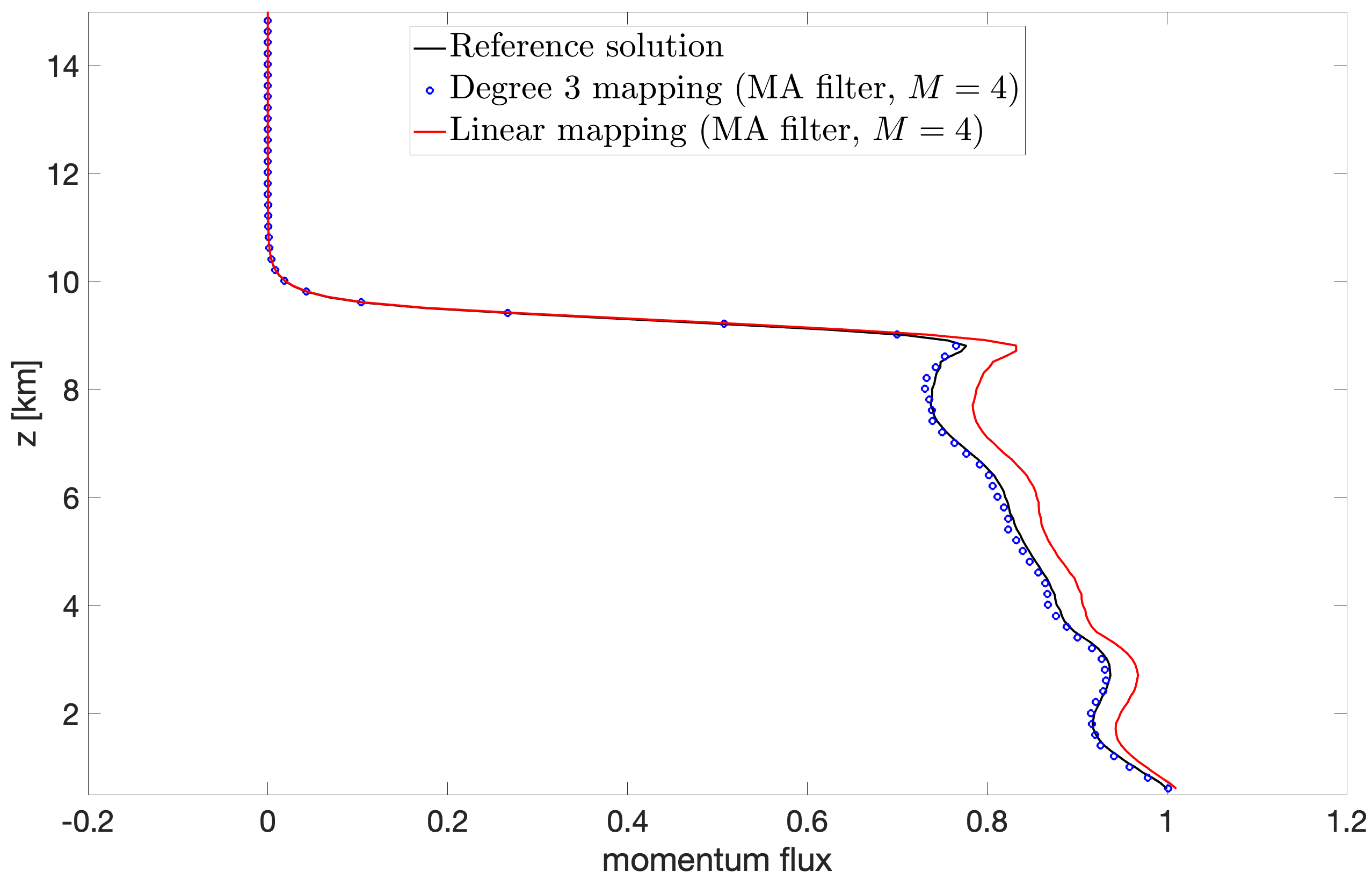}
	\end{subfigure}
	\begin{subfigure}{0.475\textwidth}
		\centering
		\includegraphics[width = 0.95\textwidth]{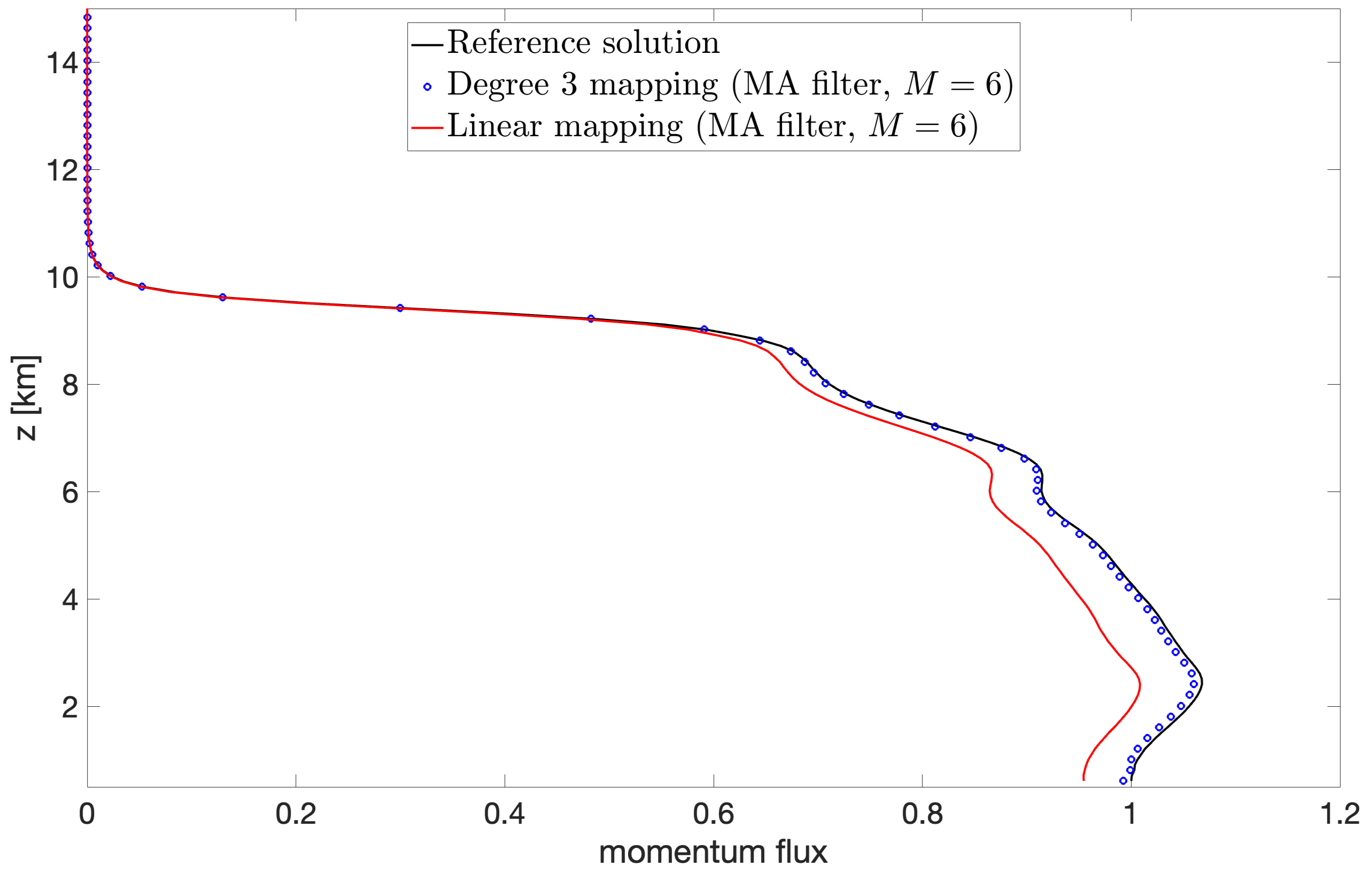}
	\end{subfigure}
	\begin{subfigure}{0.475\textwidth}
		\centering
		\includegraphics[width = 0.95\textwidth]{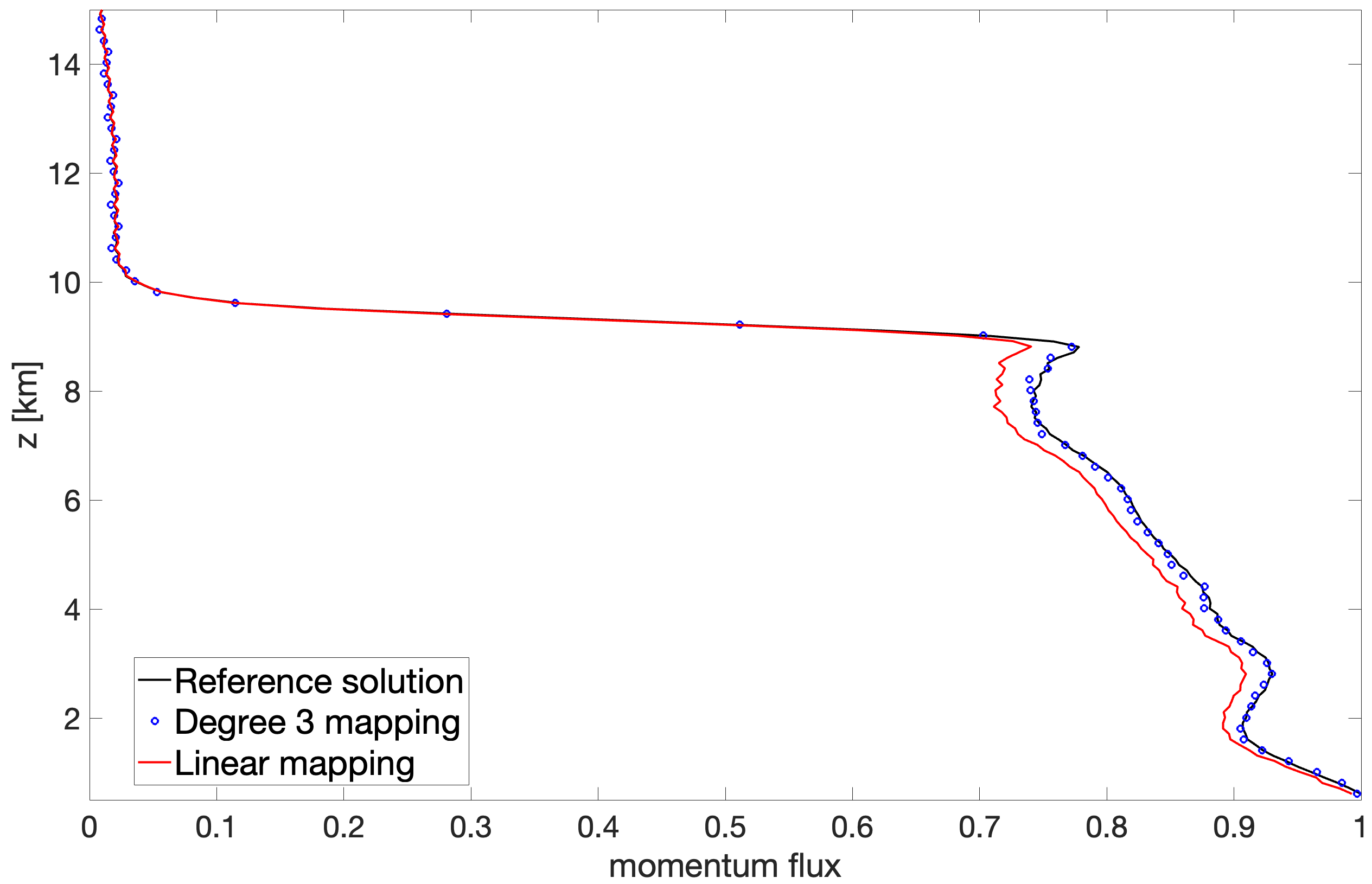}
	\end{subfigure}
	\caption{Flow over non-smooth orography with $\delta = 0.15$in \eqref{eq:non_smooth_orography} filtered using the moving-average filter \eqref{eq:ma} and $M = 4$ (top row), $M = 6$ (bottom row), normalized momentum flux computed with \eqref{eq:momentum_flux_isolated_orography} at $t = \frac{T_{f}}{2}$ (left column) and $t = T_{f}$ (right column). Results with polynomial degree 3 mapping (blue dots), linear mapping (red line), and reference results with a $300 \times 50$ elements mesh and a linear mapping (black line). The momentum flux is normalized by the value at the $z = \SI{550}{\meter}$ obtained with the $300 \times 50$ elements mesh.}
	\label{fig:non_smooth_normalized_momentum_flux_comparison_ma_M4_M6}
\end{figure}

First, we analyze the results obtained with the moving-average filter \eqref{eq:ma} (Figure \ref{fig:non_smooth_filtered_orographies_ma}). For \(M = 4\) and in terms of \(l^{2}\) relative error, a comparison of the normalized momentum flux \eqref{eq:momentum_flux_isolated_orography} shows that the high-order mapping outperforms the linear mapping by a factor 3 at \(t = \frac{T_{f}}{2}\) and by a factor 4 at \(t = T_{f}\) (Figure \ref{fig:non_smooth_normalized_momentum_flux_comparison_ma_M4_M6} and Table \ref{tab:errors_non_smooth_orography_ma}). Analogous results are obtained for \(M = 6\), for which the better accuracy established by the high-order mapping is even more evident. This is due to the fact that only few small-scale topographic features are preserved by the filter, and, at a coarse resolution, such features can be properly captured only by means of a high-order mapping. The overhead of the simulations employing the high-order mapping with respect to those using the linear mapping amounts to around $15-20\%$ in terms of wall-clock time (Table \ref{tab:errors_non_smooth_orography_ma}).

\begin{table}[h!]
	\centering
	\footnotesize
	\begin{tabularx}{0.92\columnwidth}{lcXXcXXc}
		\toprule
		\multirow{2}{*}{$M$} & \multirow{2}{*}{$t$} & \multicolumn{2}{c}{Linear mapping} & & \multicolumn{2}{c}{High-order mapping} \\
		\cmidrule(l){3-4}\cmidrule(l){6-8}
		& & error $m(z)$ \eqref{eq:momentum_flux_isolated_orography} & error $m(z)$ \eqref{eq:momentum_flux} & & error $m(z)$ \eqref{eq:momentum_flux_isolated_orography} & error $m(z)$ \eqref{eq:momentum_flux} & Overhead \\
		\midrule
		\multirow{2}{*}{4} & $\frac{T_{f}}{2} = \SI{3}{\hour}$ & $7.97 \times 10^{-2}$ & $8.04 \times 10^{-2}$ & & $2.52 \times 10^{-2}$ & $2.55 \times 10^{-2}$ & \\
		\cmidrule(l){2-8}
		& $T_{f} = \SI[parse-numbers=false]{6}{\hour}$ & $4.16 \times 10^{-2}$ & $4.05 \times 10^{-2}$ & & $7.39 \times 10^{-3}$ & $9.97 \times 10^{-3}$ & $12\%$ \\
		\midrule
		\multirow{2}{*}{6} & $\frac{T_{f}}{2} = \SI{3}{\hour}$ & $5.39 \times 10^{-2}$ & $5.42 \times 10^{-2}$ & & $6.00 \times 10^{-3}$ & $5.97 \times 10^{-3}$ & \\
		\cmidrule(l){2-8}
		& $T_{f} = \SI[parse-numbers=false]{6}{\hour}$ & $2.77 \times 10^{-2}$ & $2.69 \times 10^{-2}$ & & $2.69 \times 10^{-3}$ & $4.23 \times 10^{-3}$ & $20\%$ \\
		\bottomrule
	\end{tabularx}
	\caption{Flow over non-smooth orography with $\delta = 0.15$ in \eqref{eq:non_smooth_orography} filtered using the moving-average filter \eqref{eq:ma}, $l^{2}$ relative errors on the normalized momentum flux computed with both \eqref{eq:momentum_flux_isolated_orography} and \eqref{eq:momentum_flux}. The relative error is computed with respect to the reference solution in the region $\left[z_{1}, z_{2}\right]$, with $z_{1} = \SI{1}{\kilo\meter}$ and $z_{2} = \SI{9}{\kilo\meter}$. Here, and in the following tables and figures, $M + 1$ is the number of values employed for the moving-average filter. The overhead using the high-order mapping is computed with respect to the WT of the simulation employing the linear mapping.}
	\label{tab:errors_non_smooth_orography_ma}
\end{table}

Next, we consider the spectral filter \eqref{eq:sf}. Following \cite{webster:2003}, we take \(m = 6\) and \(\varepsilon = 1\), whereas we consider 
\[\Delta = \frac{3}{20}\Delta x, \frac{1}{5}\Delta x, \frac{1}{4}\Delta x,\] with 
\[\Delta x = \frac{L}{N - 1} = \frac{100}{600} = \frac{1}{6}.\] 
The filtered orography strongly depends on the parameter \(\Delta\). The obtained results do not show a visible discrepancy between the linear mapping and the high-order mapping and they are extremely sensitive with respect to the parameter \(\Delta\).

%%%%%%%%%%%%%%%%%%%%%%%%%%%%% T-REX %%%%%%%%%%%%%%%%%%%%%%%%%%%%%%%%%%
\subsection{T-REX Mountain-Wave}
\label{ssec:trex} \indent

\begin{figure}[h!]
	\centering
	\includegraphics[width = 0.7\textwidth]{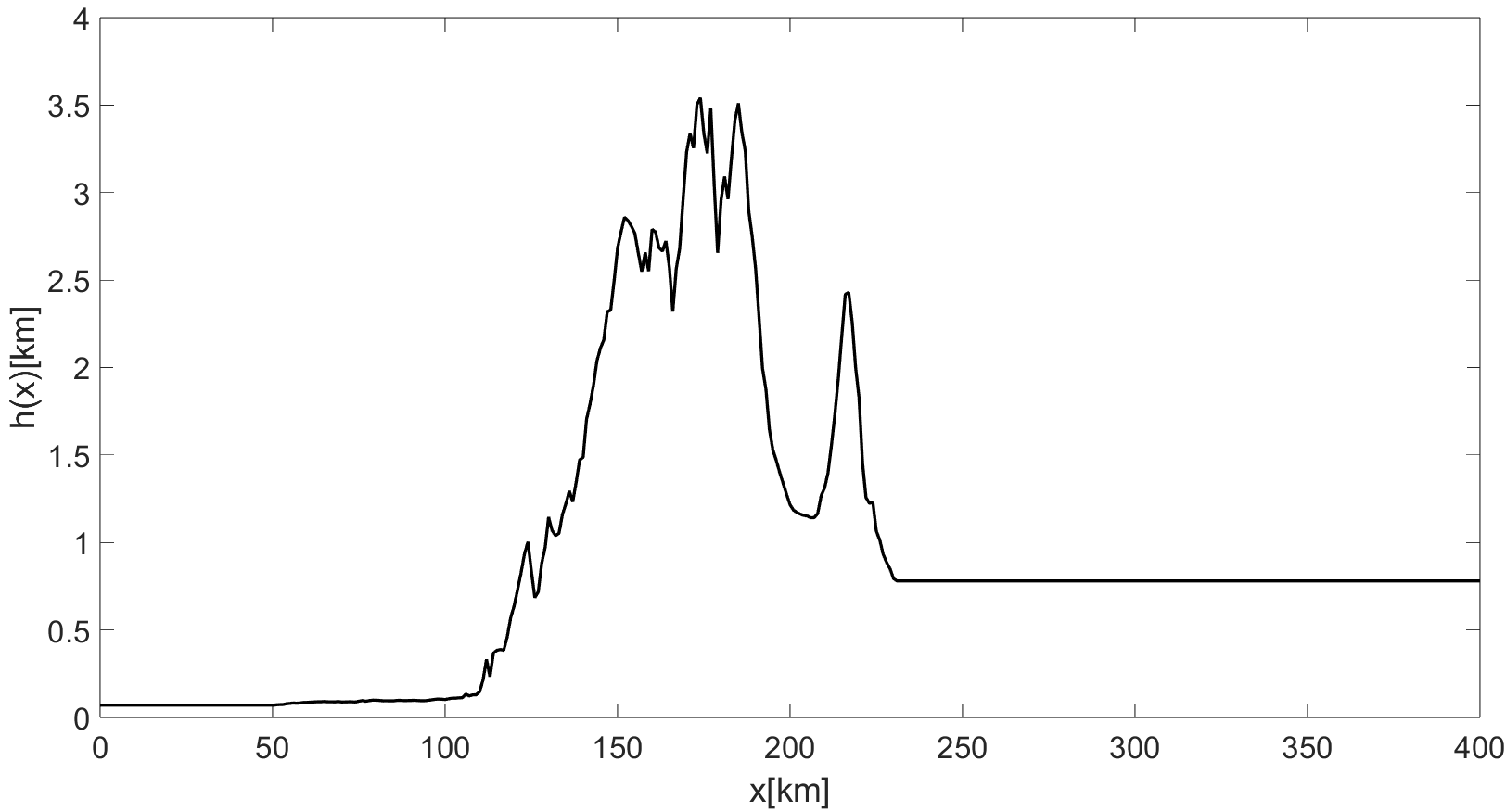}
	\caption{T-REX mountain-wave test, Sierra height profile.}
	\label{fig:TREX_profile}
\end{figure}

\begin{figure}[h!]
	\centering
	\begin{subfigure}{0.475\textwidth}
		\centering
		\includegraphics[width = 0.95\textwidth]{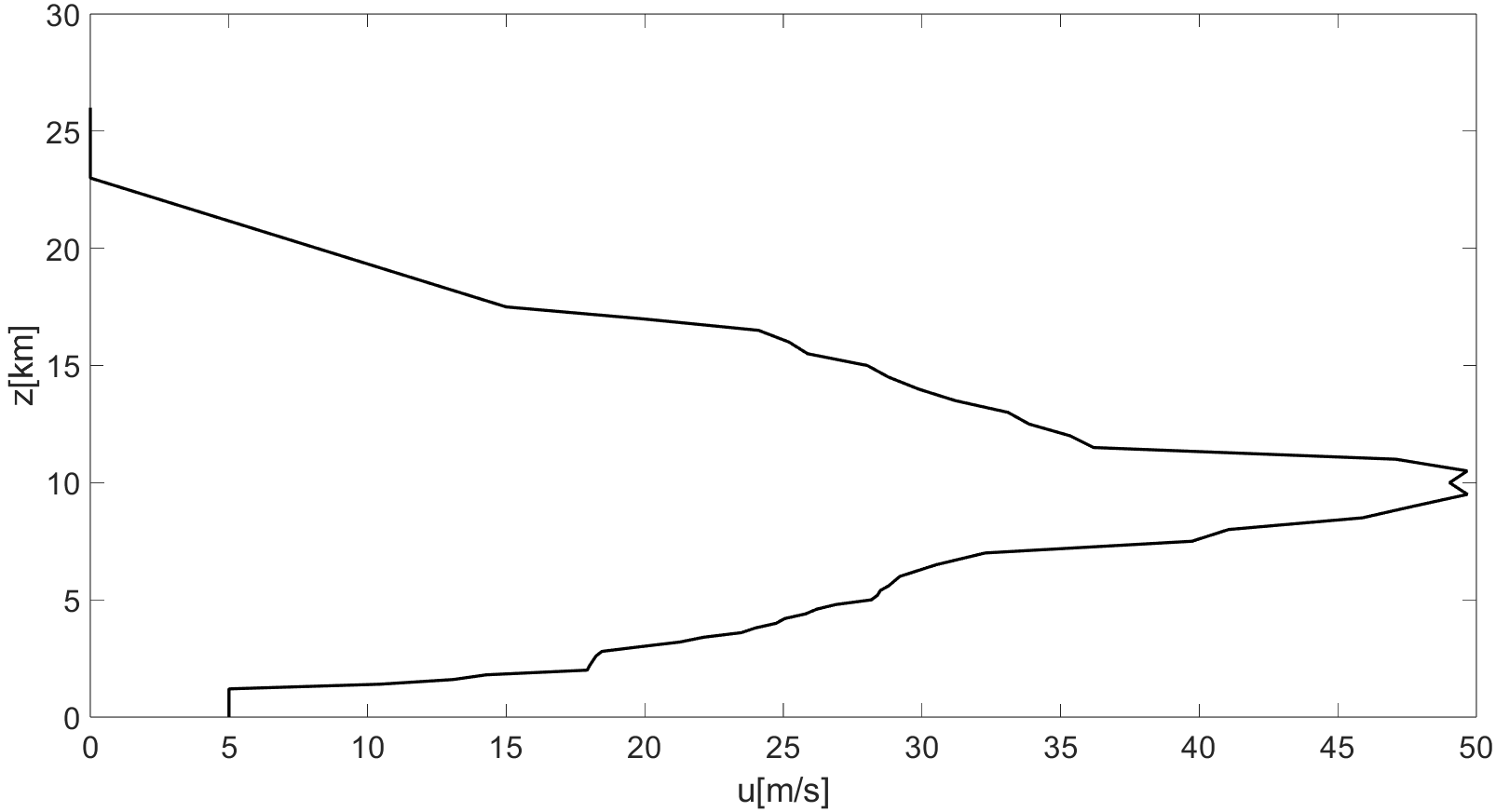}
	\end{subfigure}
	\begin{subfigure}{0.475\textwidth}
		\centering
		\includegraphics[width = 0.95\textwidth]{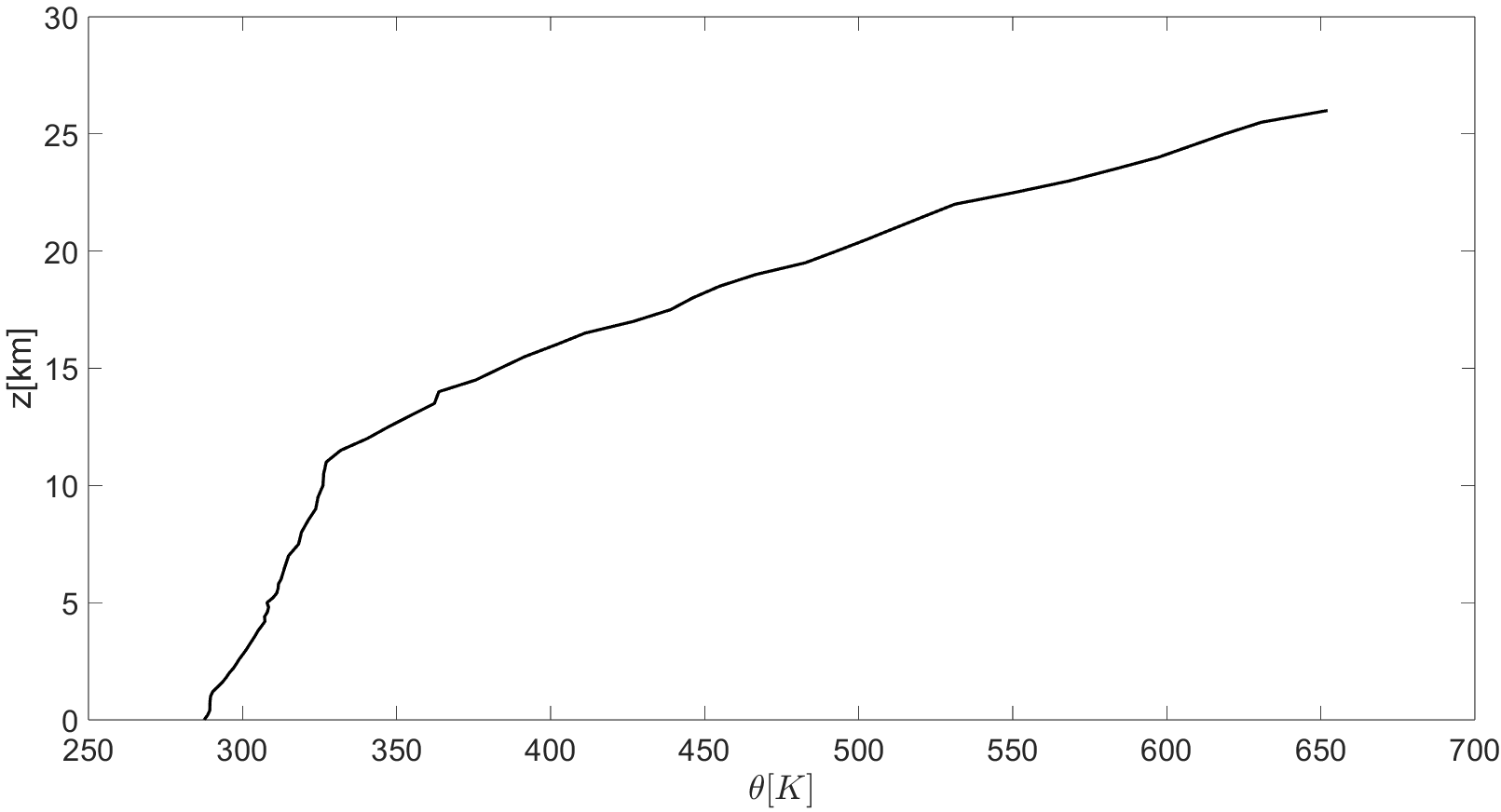}
	\end{subfigure}
	\caption{T-REX mountain-wave test case, initial conditions. Left: horizontal velocity. Right: potential temperature.}
	\label{fig:TREX_initial_conditions}
\end{figure}

In a final test, we consider simulations of a flow over a real orography profile (Figure \ref{fig:TREX_profile}, see also \cite{doyle:2011, orlando:2024}). The initial state, reported in Figure \ref{fig:TREX_initial_conditions}, is horizontally homogeneous and it is based on conditions observed during the Intensive Observation Period 6 of the Terrain-Induced Rotor Experiment (T-REX) \cite{doyle:2011}. The pressure is computed from the temperature using hydrostatic balance, namely
\begin{equation}
	p(z) = p_{0}\exp{\left(-\frac{g}{R}\int_{0}^{z} \frac{1}{T(s)}ds\right)},
\end{equation}
with \(p_{0} = \SI[parse-numbers=false]{10^{5}}{\pascal}\). Linear interpolation is used to evaluate both temperature and horizontal velocity
outside the given data points. We consider a DG spatial discretization using degree \(r = 2\) polynomials. The computational mesh is composed by \(150 \times 52\) elements, yielding a resolution of around \(\SI{133.33}{\meter}\) along the horizontal direction and of \(\SI{250}{\meter}\) along the vertical direction. A reference solution is computed using a computational mesh composed by \(400 \times 52\) elements. Note that, as reported in \cite{doyle:2011}, the orographic profile has already been filtered to remove high frequency variations. Therefore, we do not employ the filtering approaches described in Section \ref{ssec:non_smooth}. The vertical turbulent diffusion model \eqref{eq:turbulent_diffusion} is again necessary to obtain a stable numerical solution and it is used with \(l = \SI{100}{\meter}\) and \(\theta_{0} = \SI{273}{\kelvin}\).

\begin{figure}[h!]
	\centering
	\begin{subfigure}{0.9\textwidth}
		\centering
		\includegraphics[width = 0.7\textwidth]{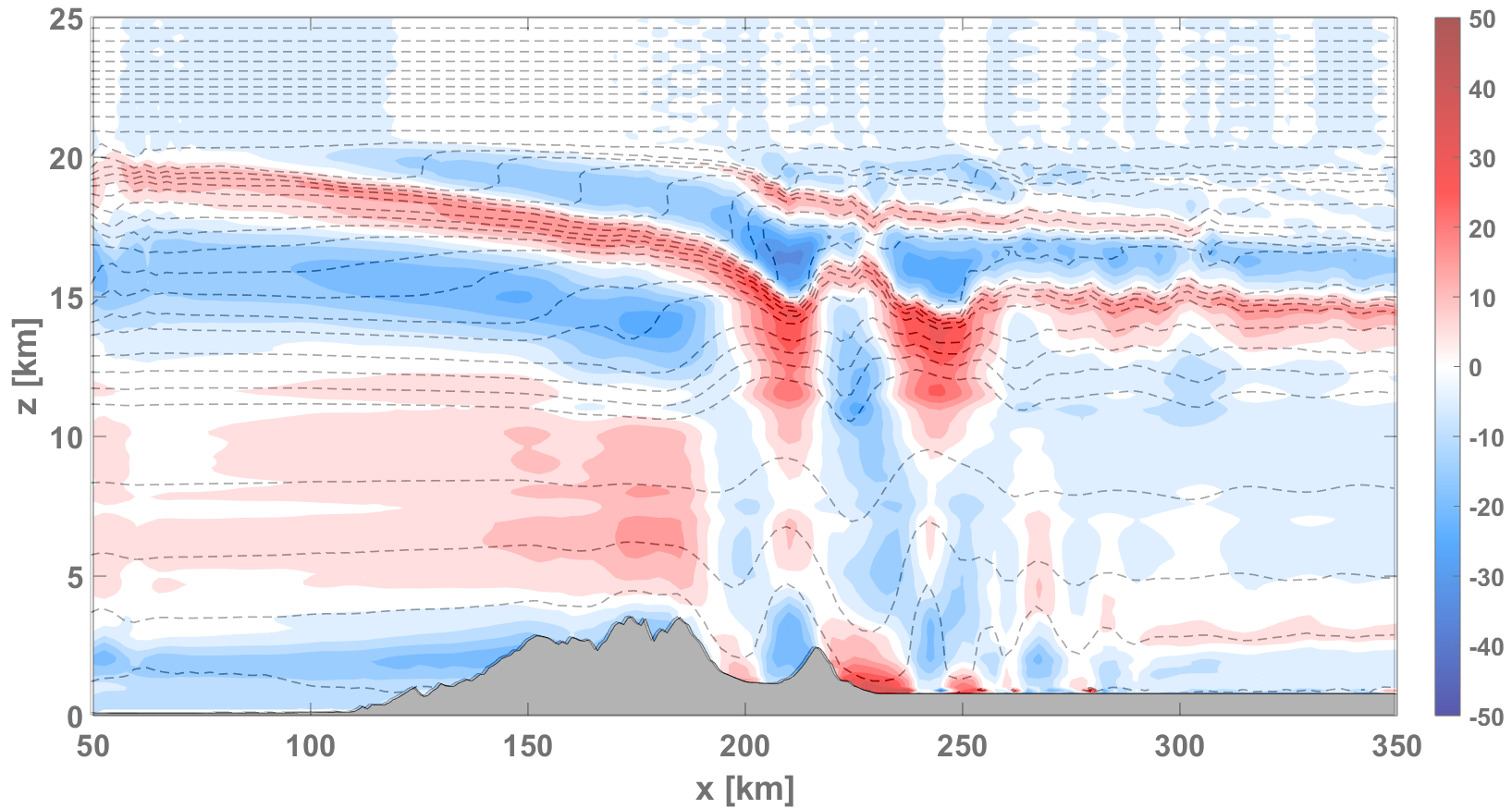}
	\end{subfigure}
	\begin{subfigure}{0.9\textwidth}
		\centering
		\includegraphics[width = 0.7\textwidth]{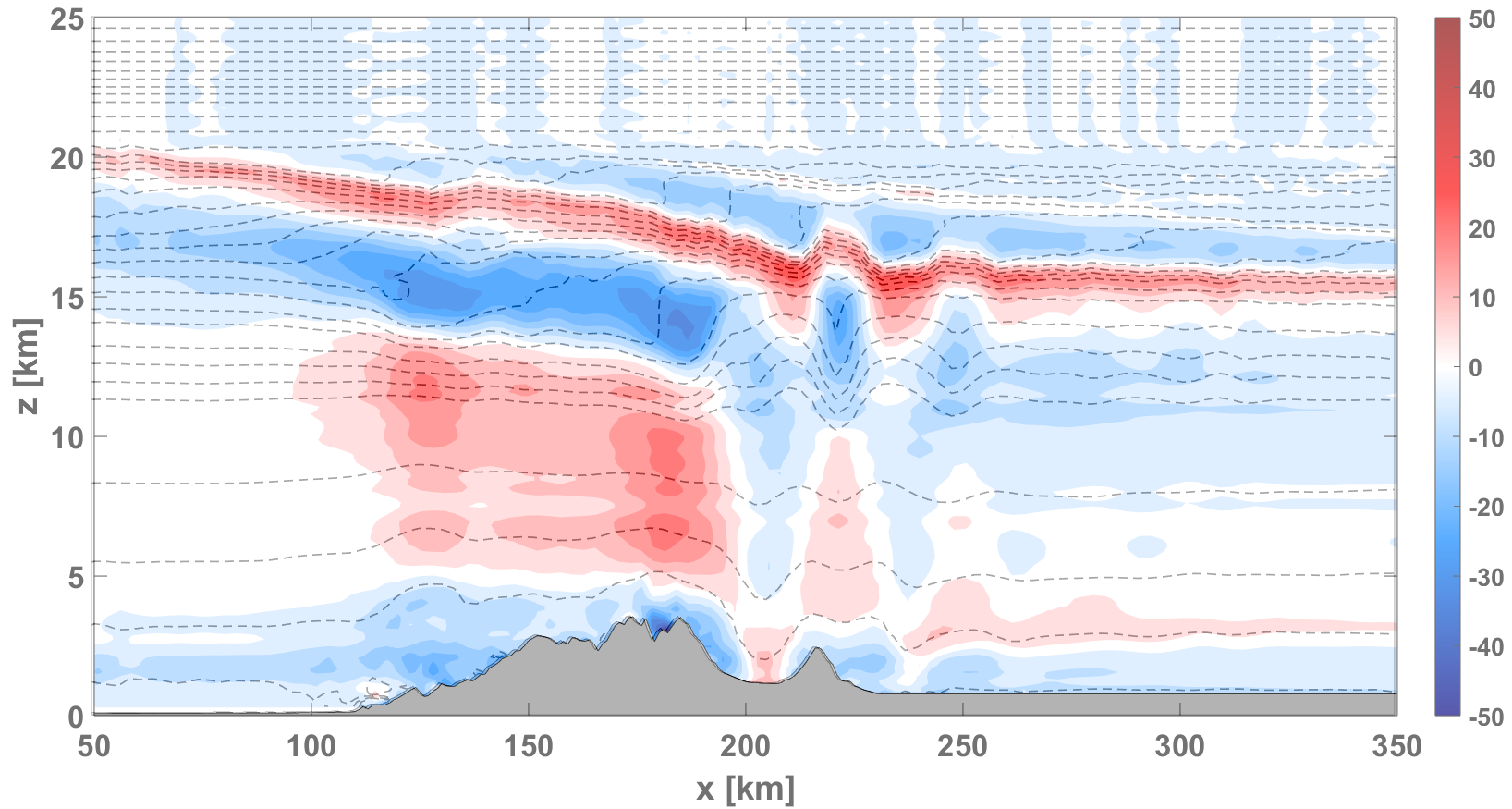}
	\end{subfigure}
	\begin{subfigure}{0.9\textwidth}
		\centering
		\includegraphics[width = 0.7\textwidth]{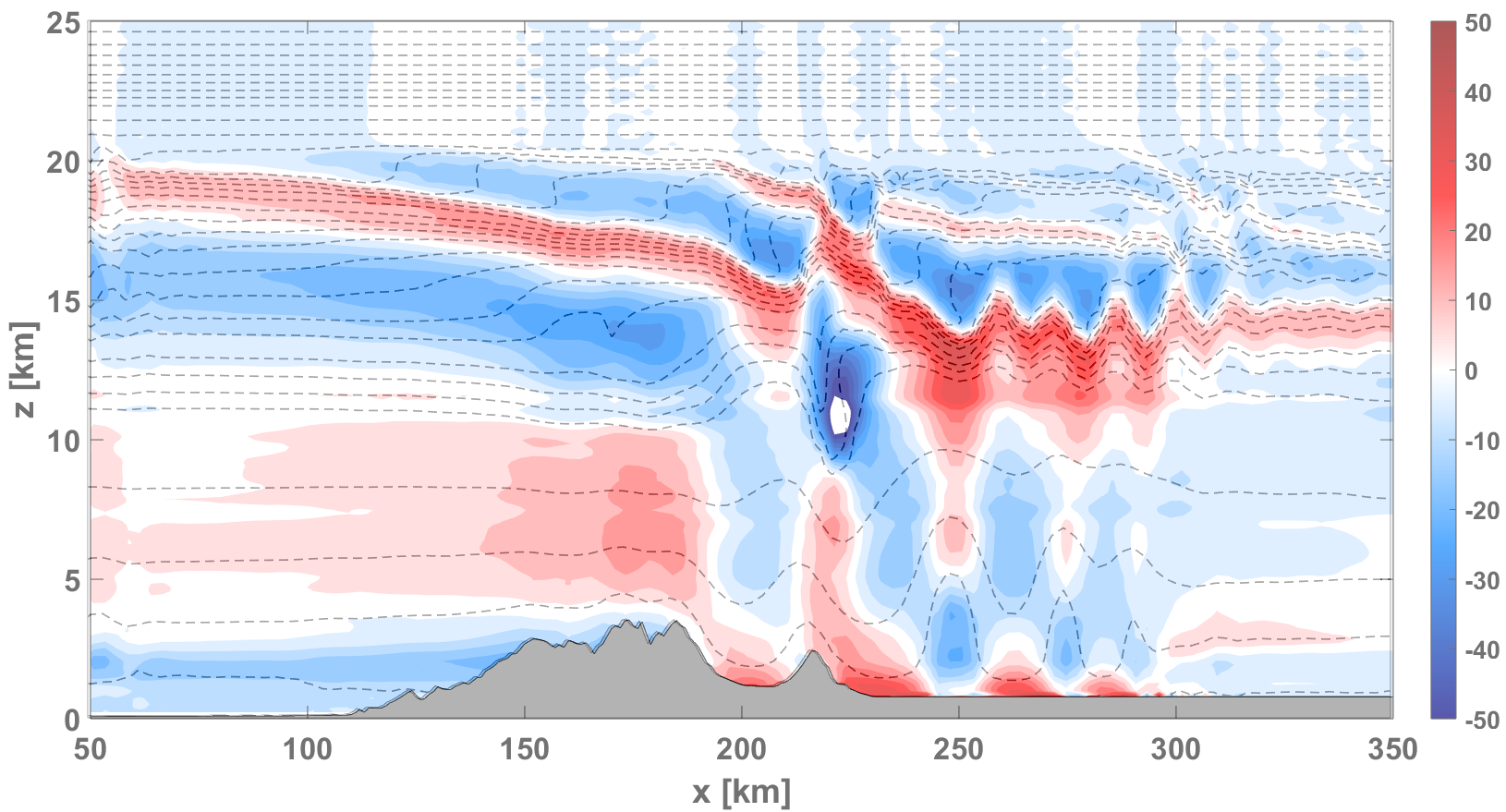}
	\end{subfigure}
	\caption{T-REX mountain wave test case at $t = T_{f} = \SI{4}{\hour}$. Top: Reference solution. Middle: Polynomial degree $3$ mapping. Bottom: Linear mapping. Horizontal velocity perturbation (colors), contours in the range $\SI[parse-numbers=false]{[-50,50]}{\meter\per\second}$ with a $\SI{5}{\meter\per\second}$ interval. Potential temperature (dashed lines), contours in the range $\SI[parse-numbers=false]{[270, 800]}{\kelvin}$ with a $\SI{10}{\kelvin}$ interval.}
	\label{fig:TREX_contours}
\end{figure}

A comparison of the contours of the horizontal velocity deviation and of the potential temperature shows that high-order mapping results are more similar to the reference results with respect to those obtained with the linear mapping (Figure \ref{fig:TREX_contours}). 

\begin{figure}[h!]
	\centering
	\includegraphics[width = 0.7\textwidth]{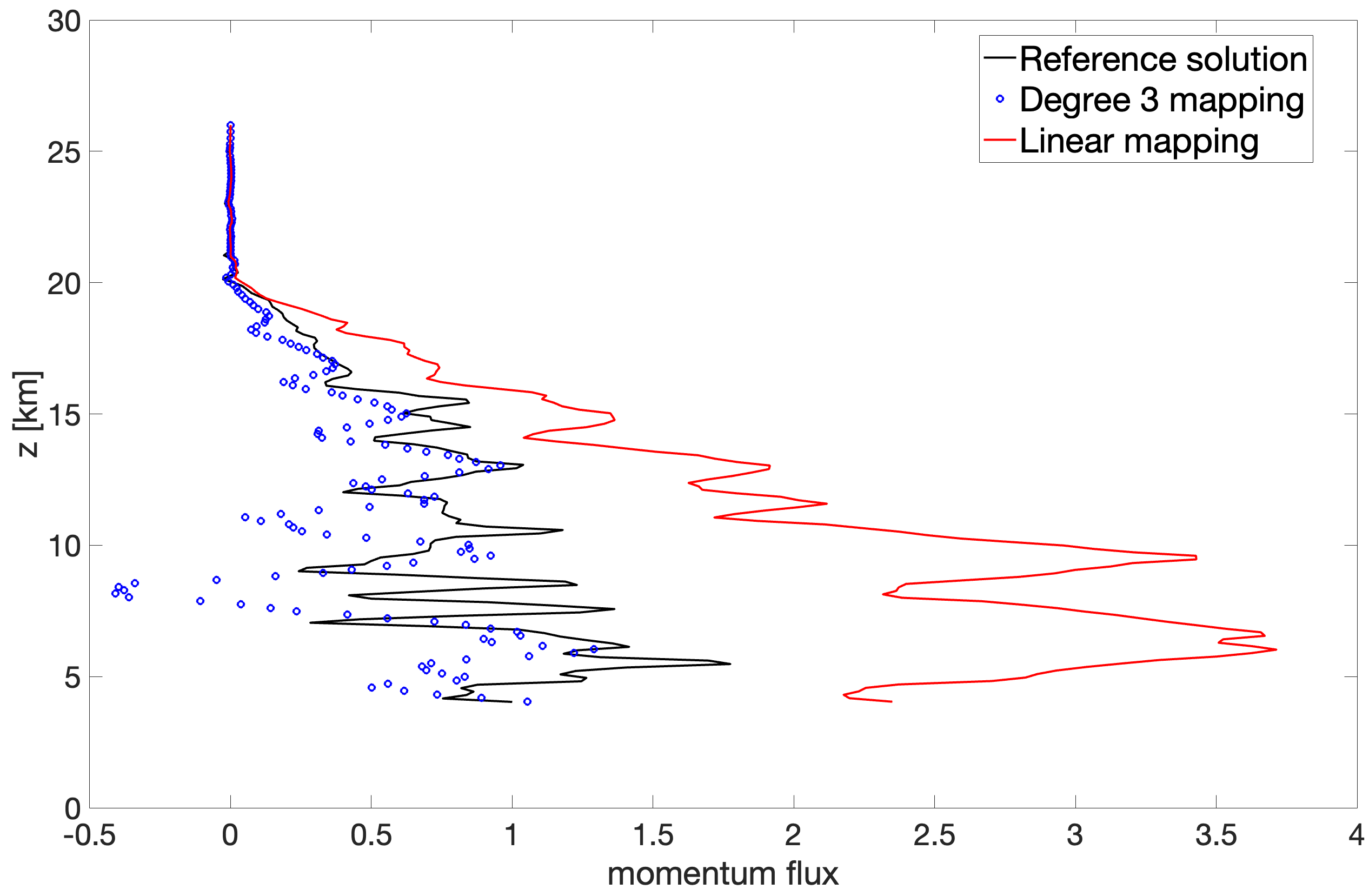}
	\caption{T-REX mountain wave test case, comparison of normalized momentum flux at $t = T_{f} = \SI{4}{\hour}$ computed with \eqref{eq:momentum_flux}. Results using polynomial degree 3 mapping (blue dots), linear mapping (red line), and reference results with a $400 \times 52$ elements mesh and a linear mapping (black line). The momentum flux is normalized by the value at $z = \SI{5}{\kilo\meter}$ obtained with the $400 \times 52$ elements mesh.}
	\label{fig:TREX_normalized_momentum_flux_comparison}
\end{figure}

Finally, we provide a more quantitative comparison by computing the vertical momentum flux. Following \cite{doyle:2011}, we use the average value of the velocity along the horizontal and vertical direction to compute \(u'\) and \(w'\) in \eqref{eq:momentum_flux_isolated_orography}. As discussed in \cite{doyle:2011}, there is low predictability of key characteristics such as the strength of downslope winds or the location and intensity of stratospheric wave breaking. Moreover, the change in the resolution of the orography has been shown to modify the representation of mountain wave-driven middle atmosphere processes \cite{kanehama:2019}. Nevertheless, the values of the momentum flux obtained with the high-order mapping are much closer to the reference results than those obtained with the linear mapping, see Figure \ref{fig:TREX_normalized_momentum_flux_comparison} and Table \ref{tab:errors_TREX}. In terms of \(l^{2}\) relative error, the high-order mapping is almost three times smaller than the error using the linear mapping. These results corroborate those obtained in the non-smooth orography test case, and confirm the advantage of high-order mapping over linear mapping in resolving small-scale orographic features.
The overhead of the simulations employing the high-order mapping with respect to those using the linear mapping amounts to around the $37\%$ in terms of wall-clock time (Table \ref{tab:errors_TREX}).

\begin{table}[h!]
	\centering
	\footnotesize
	\begin{tabularx}{0.92\columnwidth}{XXcXXc}
		\toprule
		\multicolumn{2}{c}{Linear mapping} & & \multicolumn{2}{c}{High-order mapping} \\
		\cmidrule(l){1-2}\cmidrule(l){4-6}
		error $m(z)$ \eqref{eq:momentum_flux_isolated_orography} & error $m(z)$ \eqref{eq:momentum_flux} & & error $m(z)$ \eqref{eq:momentum_flux_isolated_orography} & error $m(z)$ \eqref{eq:momentum_flux} & Overhead \\
		\midrule
		$1.64$ & $1.84$ & & $4.42 \times 10^{-1}$ & $5.85 \times 10^{-1}$ & $37\%$ \\
		\bottomrule
	\end{tabularx}
	\caption{T-REX mountain wave test case, $l^{2}$ relative errors on the normalized momentum flux at $t = T_{f} = \SI{4}{\hour}$ computed with both \eqref{eq:momentum_flux_isolated_orography} and \eqref{eq:momentum_flux}. The relative error is computed with respect to the reference solution in the region $\left[z_{1}, z_{2}\right]$, with $z_{1} = \SI{5}{\kilo\meter}$ and $z_{2} = \SI{20}{\kilo\meter}$. The overhead using the high-order mapping is computed with respect to the WT time of the simulation employing the linear mapping.}
	\label{tab:errors_TREX}
\end{table}

%%%%%%%%%%%%%%%%%%%% Comment on computational cost %%%%%%%%%%%%%%%%
\subsection{Considerations on the computational cost}
\label{ssec:WT_times}
	
In this Section, we report some considerations on the computational cost. The use of high-order mappings has an impact on the quadrature rule \eqref{eq:quadrature_formula}. More specifically, the determinant of the Jacobian is not a constant, but a polynomial. In this work, following, e.g., \cite{giraldo:2006}, we consider exact integration. In order to achieve this goal, we employ the so-called over-integration or consistent integration. This means that we use more than \(r + 1\) Gauss quadrature points along each coordinate direction and, in particular, we use \(2r + 1\) quadrature points in the case of linear mapping and \(2r + 1 + \lceil \hat{r} - 2\rceil\) quadrature points in the case of high-order mapping, where \(\hat{r}\) is the polynomial degree of the high-order mapping. We also recall that \(r\) denotes the polynomial degree employed for the spatial discretization. One can easily notice that, for \(\hat{r} = 2\), the same number of quadrature points employed for linear mapping are sufficient to obtain exact integration. The overhead in the use of high-order mapping is therefore due to the use of higher order quadrature formulas for \(\hat{r} > 2\). 
	
However, one can notice from the results in Section \ref{ssec:hydrostatic}-\ref{ssec:trex} that the overhead typically amounts only to around $\approx 5-10\%$ and, for the test cases in Section \ref{ssec:hydrostatic} and Section \ref{ssec:schar}, the simulations using high-order mappings can be even faster. This is likely due to the fact that the use of higher order quadrature formulas is limited to the curved boundary, namely on a very small number of faces. Moreover, an efficient implementation of quadrature rules in the \texttt{deal.II} library helps to avoid overhead when high-order mappings are employed. More specifically, a variable $\mathit{JxW}$, which stands for ``Jacobian determinant times weight'', is stored. The two factors, i.e. the determinant of the Jacobian and the weights of the quadrature formula, are always coupled and, therefore, storing only their product allows an efficient evaluation of integrals and saves computational operations. We refer to \cite{bangerth:2007} for further details on the specific implementation in the library. Hence, for \(\hat{r} = 2\), the computational cost is the same of the linear mapping (see Table \ref{tab:overhead_linear_hydro}), while for \(\hat{r} > 2\) the overhead in terms of wall-clock time is reduced and, because of caching effects, the computational cost can be even lower.
	
The only test cases in which a sizeable overhead appears are the unfiltered non-smooth topography with perturbation factor $\delta = 0.15$ in \eqref{eq:non_smooth_orography} and, with a reduced overhead, the T-REX mountain wave (Tables \ref{tab:errors_non_smooth_orography_delta0,15} and \ref{tab:errors_TREX}). This result is likely due to the fact that the high-order mapping captures and resolves small-scale flow patterns that lead to a more complex flow field and strongly affect the development of lee waves (see Figure \ref{fig:non_smooth_normalized_momentum_flux_comparison_delta0,15}). This is confirmed by the fact that the overhead factor is significantly reduced employing a filtered orography. These considerations further support the use of high-order mappings, which significantly outperform linear mapping in terms of accuracy with an increase in computational cost that only depends on the more complex flow features induced by the better resolved orography. However, the wall-clock time needed for the high-order mapping is more than twice smaller than the wall-clock needed for the reference simulation employing the linear mapping.

%%%%%%%%%%%%%%%%%%%%%%%%%%%%%%%%% Conclusions %%%%%%%%%%%%%%%%%%%%%%%%%
\section{Conclusions}
\label{sec:conclu} \indent

The accurate resolution of orographic profiles is of paramount importance for atmospheric applications \cite{orlando:2024, prusa:2006}. Here we have performed a quantitative study on the impact of curved elements for flows over orography. To the best of our knowledge, this study has never been performed for atmospheric applications. We have employed the IMEX-DG solver originally proposed in \cite{orlando:2022} and validated for atmospheric applications in \cite{orlando:2023a, orlando:2024}. The software, implemented in the framework of the open-source library \texttt{deal.II} \cite{arndt:2023}, natively supports high-order polynomials for the mapping between the reference element and elements in the physical space. 

Numerical experiments on a number of classical benchmarks of flow over idealised and real orography have shown that, at a given resolution, results obtained with a high-order mapping significantly outperform those obtained using a linear mapping in terms of error with respect to reference solutions. A sizeable increase in computational cost is only observed when more complex flow features are induced by the better resolved orography. In fact, the use of high-order mapping leads to results which are analogous to those obtained using a linear mapping at orography-resolving high resolution. Hence, employing high-order mapping is to some extent equivalent to considering a sub-tessellation to evaluate the integrals. Applications of the use of high-order mappings to non-smooth orography profiles have also been presented, showing that the use of high-order mappings generally provides better results with respect to those obtained with a linear mapping. This is valid also for orography profiles obtained with standard filtering procedures. At a given spatial resolution, high-order mappings capture small-scale topographic features inaccessible to the standard linear mapping. Performing very high resolution simulations is computationally expensive, but it is beneficial for mountain atmospheric processes and forecast skill. The use of high-order mappings can be considered as a valuable alternative tool to resolve orographic features at lower spatial resolutions and at a feasible computational cost. Furthermore, based on the numerical evidence presented in this paper, we believe that only results taking into account improved representations of the geometry should be considered as the reference for future works.

%%%%%%%%%%%%%%%%%%%%%%%%% Acknowledgements %%%%%%%%%%%%%%%%%%%%%%%%%%
\section*{Acknowledgements}

We thank the Associate Editor Dr. P. Smolarkiewicz and an anonymous reviewer for their comments and suggestions which helped to improve the quality of the paper. G.O. is part of the INdAM-GNCS National Research Group. The simulations have been partly run at CINECA thanks to the computational resources made available through the ISCRA-C projects FEMTUF - HP10CTQ8X7 and FEM-GPU - HP10CQYKJ1. We acknowledge the CINECA award, for the availability of high-performance computing resources and support. This work has been partly supported by the ESCAPE-2 project, European Union’s Horizon 2020 Research and Innovation Programme (Grant Agreement No. 800897). The authors also gratefully acknowledge Dr. Doyle and Dr. Gaber{\u{s}}ek for having provided the orography profile and the initial data of the T-REX mountain wave.

\bibliographystyle{cas-model2-names}
\bibliography{DG_Orography_Curved_Boundary.bib}
	
\end{document}